\documentclass[a4paper,fleqn]{cas-dc}

\usepackage[numbers]{natbib}

\graphicspath{{figures/}}

\usepackage{algorithm}
\usepackage{algorithmic}

\makeatletter
\renewcommand\printorcid{}
\makeatother

\providecommand{\textproc}[1]{\textsc{#1}}

\newcommand{\vAbstUnavoidableOverlap}{89}

\newcommand{\vAbstUnsatOverlap}{21.4}
\newcommand{\vAbstUnsatPSNROverlap}{20.9}
\newcommand{\vAbstUnsatPSNRRecency}{19.8}

\newcommand{\vAbstUsableOverlap}{0.5}
\newcommand{\vAbstUsableRecency}{12.6}
\newcommand{\vAffineFPMAE}{5.42}
\newcommand{\vAlignMaxMs}{62.6}
\newcommand{\vAlignMedianMs}{20.3}
\newcommand{\vAlignPninetyfiveMs}{46.3}
\newcommand{\vBlockSweepMax}{31.86}
\newcommand{\vBlockSweepMin}{31.64}
\newcommand{\vBoundedSearchCost}{0.006}
\newcommand{\vCeilingCost}{0.06}
\newcommand{\vCeilingSavingMs}{6}
\newcommand{\vCeiling}{35.34}
\newcommand{\vCfgBlockRadius}{3}
\newcommand{\vCfgBlockScale}{0.35}
\newcommand{\vCfgBlock}{48}
\newcommand{\vCfgCore}{0.65}
\newcommand{\vCfgRegIterations}{15}
\newcommand{\vCfgRegScale}{0.10}
\newcommand{\vCfgResidualBand}{8}
\newcommand{\vCfgResidualRadius}{5}
\newcommand{\vCfgRoiPad}{24}
\newcommand{\vCfgSegConfidence}{0.25}
\newcommand{\vCfgSubpixelSteps}{4}
\newcommand{\vCfgSupport}{0.65}
\newcommand{\vCfgUncertain}{0.35}
\newcommand{\vCleanOverlap}{33.37}
\newcommand{\vCleanRecency}{33.42}
\newcommand{\vFixedCohort}{84}
\newcommand{\vHeldGainClipHigh}{+2.20}
\newcommand{\vHeldGainClipLow}{+1.04}
\newcommand{\vHeldGainVideoFirst}{+1.96}
\newcommand{\vHeldGainVideoHigh}{+1.99}
\newcommand{\vHeldGainVideoLow}{+1.32}
\newcommand{\vHeldGainVideoSecond}{+1.32}
\newcommand{\vHeldGainVideoThird}{+1.99}
\newcommand{\vHeldGain}{1.58}
\newcommand{\vHeldInstrumentCoverage}{97.3}
\newcommand{\vHeldIoU}{0.744}
\newcommand{\vHeldPSNR}{29.83}
\newcommand{\vHeldPninetyfive}{86.1}
\newcommand{\vHeldPrecision}{0.757}
\newcommand{\vHeldRecall}{0.973}
\newcommand{\vHeldSSIM}{0.893}
\newcommand{\vHeldSamplesVideoFirst}{36}
\newcommand{\vHeldSamplesVideoSecond}{60}
\newcommand{\vHeldSamplesVideoThird}{5}

\newcommand{\vLagOneFixed}{33.37}
\newcommand{\vLagOne}{33.42}
\newcommand{\vLagTwelveDelta}{-5.31}
\newcommand{\vLagTwelve}{28.06}
\newcommand{\vLagTwoDelta}{-1.43}
\newcommand{\vLamaHeldPSNR}{30.17}

\newcommand{\vLamaMatchedGainHeld}{3.77}
\newcommand{\vLamaMatchedGain}{3.45}

\newcommand{\vLamaValPSNR}{31.89}

\newcommand{\vNavierStokes}{28.58}
\newcommand{\vOccAbstainOverlap}{3.7}
\newcommand{\vOccAbstainRecency}{15.4}
\newcommand{\vOccGainHeavy}{4.40}
\newcommand{\vOccGainLight}{1.34}
\newcommand{\vOccGainOverPublished}{3.63}
\newcommand{\vOccGain}{3.60}

\newcommand{\vOccMinSSD}{26.41}
\newcommand{\vOccOverlapOverRanked}{2.90}
\newcommand{\vOccOverlap}{30.04}
\newcommand{\vOccRankedGain}{0.70}
\newcommand{\vOccRanked}{27.14}
\newcommand{\vOccRecency}{26.44}
\newcommand{\vOursHeldTruthPSNR}{33.94}

\newcommand{\vOursTruthPSNR}{35.34}

\newcommand{\vPasteGainOverInpainting}{1.68}
\newcommand{\vPaste}{30.26}
\newcommand{\vPreBoundPninetyfive}{96.5}
\newcommand{\vPreBoundPninetynine}{115.4}
\newcommand{\vQualityControlPSNR}{33.44}
\newcommand{\vQualityLatencyMean}{55.8}
\newcommand{\vQualityPSNR}{33.43}
\newcommand{\vQualityPninetyfive}{86.2}
\newcommand{\vResidualTranslationGain}{1.55}
\newcommand{\vSegFOne}{0.923}
\newcommand{\vSegMapCoarse}{0.756}
\newcommand{\vSegMedianMs}{13.8}
\newcommand{\vSegPninetyfiveMs}{14.5}
\newcommand{\vSegPrecision}{0.962}
\newcommand{\vSegRecall}{0.886}
\newcommand{\vSegmentationGap}{3.45}
\newcommand{\vSelectedFPMAE}{3.96}
\newcommand{\vSelectedInstrumentCoverage}{97.6}
\newcommand{\vSelectedIoU}{0.769}

\newcommand{\vSelectedOutside}{0.010}
\newcommand{\vSelectedPSNR}{31.89}
\newcommand{\vSelectedPfifty}{51.7}
\newcommand{\vSelectedPninetyfive}{85.5}
\newcommand{\vSelectedPninetynine}{101.4}
\newcommand{\vSelectedPrecision}{0.783}
\newcommand{\vSelectedRecall}{0.976}
\newcommand{\vSelectedSSIM}{0.915}
\newcommand{\vSelectionSpread}{4.23}
\newcommand{\vSigCleanIdentical}{173}
\newcommand{\vSigDeployedBndP}{0.22}
\newcommand{\vSigDeployedPSNRP}{0.34}
\newcommand{\vSigMatchedBndWins}{139}
\newcommand{\vSigMatchedMAEWins}{145}
\newcommand{\vSigMatchedPSNRWins}{135}
\newcommand{\vSigMatchedSSIMWins}{141}
\newcommand{\vSigN}{178}
\newcommand{\vSigOccDiffering}{78}
\newcommand{\vSigOccIdentical}{100}
\newcommand{\vSigOccOverlapWins}{66}
\newcommand{\vSigPubDiffering}{84}
\newcommand{\vSigPubOverlapWins}{71}
\newcommand{\vTelea}{28.47}
\newcommand{\vTransrefHeldPSNR}{27.54}

\newcommand{\vTransrefMatchedGainHeld}{6.40}
\newcommand{\vTransrefMatchedGain}{5.54}
\newcommand{\vTransrefValPSNR}{29.80}

\makeatletter
\AtEndDocument{\clearpage\addtocounter{page}{-1}%
  \immediate\write\@auxout{\string\csxdef{lastpage}{\thepage}}%
  \addtocounter{page}{1}}
\makeatother

\begin{document}
\let\WriteBookmarks\relax
\def\floatpagepagefraction{1}
\def\textpagefraction{.001}

\shorttitle{TransGHOST: Real-Time, Synthesis-Free Reconstruction of Instrument-Occluded Regions in Ophthalmic Microsurgery}

\shortauthors{N. Nataraj, M. Sogabe et al.}

\title [mode = title]{TransGHOST: Real-Time, Synthesis-Free Reconstruction of Instrument-Occluded Regions in Ophthalmic Microsurgery}

\author[1]{Niran Nataraj}
\ead{niran@g.ecc.u-tokyo.ac.jp, hi@niran.dev}

\author[2]{Maina Sogabe}
\cormark[1]
\ead{fsogabe-vet@g.ecc.u-tokyo.ac.jp}

\author[3]{Norihiko Ito}
\ead{taromobile@me.com}

\author[1]{Kenji Kawashima}
\ead{kkawa729@g.ecc.u-tokyo.ac.jp}

\affiliation[1]{organization={Department of Information Physics and Computing, Graduate School of Information Science and Technology, The University of Tokyo},
            city={Tokyo},
            country={Japan}}

\affiliation[2]{organization={Department of Precision Engineering, Graduate School of Engineering, The University of Tokyo},
            city={Tokyo},
            country={Japan}}

\affiliation[3]{organization={Department of Ophthalmology, Tokyo Medical University},
            city={Tokyo},
            country={Japan}}

\cortext[1]{Corresponding author}

\begin{abstract}
In robot-assisted microsurgery, precise manipulation depends on continuous visual information about the tissue surrounding and underlying the surgical instrument. Yet the instrument itself occludes this region during manipulation, creating a temporary blind area in the visual feedback available to the surgeon and robotic system. Such occlusion can constrain downstream perception, guidance, and control, particularly in ophthalmic microsurgery, where manipulation occurs within a millimetre-scale workspace under high magnification. We therefore treat instrument occlusion as a real-time robotic perception problem and present TransGHOST, a causal reference-guided reconstruction pipeline whose non-generative safe mode operates in real time. Unlike generative inpainting, it synthesises no content: each restored value is resampled from a recorded frame using geometric transforms fitted only to visible regions. The system provides per-pixel confidence and provenance and marks unsupported pixels as abstentions. We evaluate TransGHOST on a new veterinary clinical corpus of canine cataract surgery with pixel-level instrument and anatomy labels. Only the occluded region is scored, and each scoring target is verified to differ from its reference. Safe mode reaches \vSelectedPSNR~dB masked PSNR and \vSelectedSSIM{} masked SSIM at \vSelectedPfifty~ms median and \vSelectedPninetyfive~ms 95th-percentile latency. On a pre-declared held-out cohort with training-excluded instrument cutouts, it reaches \vHeldPSNR~dB masked PSNR, while matched-mask evaluation on validation shows a \vLamaMatchedGain~dB advantage over LaMa. Region-aware reference selection improves masked PSNR by \vOccGain~dB over recency when candidate references can themselves be occluded. Reference-age analysis defines the operating interval for the brief, repeated occlusions targeted by TransGHOST.
\end{abstract}


\maketitle

\section{Introduction}
\label{sec:intro}

During ophthalmic microsurgery, the surgeon views the operative field through a high-magnification live video feed \cite{Ribeiro2022Advances}. At this scale, the view must remain continuous and geometrically stable: displacements of tens of micrometres matter, and temporary concealment of a microscopic feature can leave pathology unaddressed. Surgical instruments must nevertheless enter the field and thereby occlude the tissue under treatment. Occlusion is particularly consequential during fine dissection and instrument exchange, when no alternative view of the covered tissue is available and the surgeon's cognitive load and task difficulty increase \cite{Eckardt2016, Mukash2025}. The surgeon may consequently pause or reposition the instrument, responses that prolong operating time, and any inference about the covered tissue may itself be inaccurate \cite{Way2003}.

\begin{figure}[pos=t]
  \centering
  \includegraphics{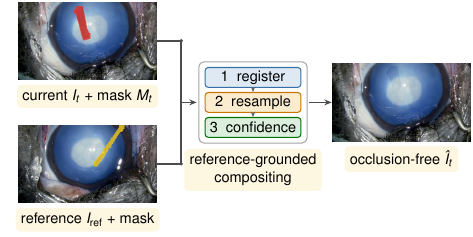}
  \caption{
    TransGHOST reconstructs the masked region from an aligned causal reference. Safe mode uses geometric resampling and returns confidence, abstention and provenance layers.
  }
  \label{fig:ref_inpaint_pipeline}
\end{figure}

Cataract surgery is the setting examined here (Figure~\ref{fig:ref_inpaint_pipeline}). Instruments such as a phacoemulsification tip and a hook are introduced into the anterior chamber through corneal incisions in order to sculpt, divide and rotate the lens nucleus \cite{Morita2020}. Each incision acts as a fulcrum that constrains the motion of the instrument passing through it, and the whole working field spans only a few millimetres. In a field of that size an instrument covering the tissue under treatment is difficult to avoid, and the surgeon commonly continues the manoeuvre rather than pausing, estimating the covered region from the frames seen a moment earlier and from the motion of the surrounding tissue.

Moving the instrument aside is not a free alternative. Its range of motion is limited by the position of the incision and by the geometry of the eye, and withdrawing and reinserting an instrument in order to look interrupts the manoeuvre and disturbs the anterior chamber and the incision itself. Presenting the tissue behind the instrument without interrupting the manipulation would therefore supply visual information that is otherwise unavailable.

The same loss of visual information is also critical for robot-assisted manipulation, because perception and control components must operate on a scene partially occluded by the instrument performing the task. Robotic systems for intraocular surgery impose the same fulcrum constraint through a remote centre of motion at the incision \cite{Iordachita2022}, so a robot-held instrument occludes the tissue in the same way. In robot-assisted ophthalmic surgery, intraoperative scene perception is built on the microscope video stream, and an instrument crossing the field removes part of the scene from any component that performs registration, tissue tracking, workflow monitoring, visual guidance or visual servoing on that stream. These components may then have to operate on incomplete visual information, potentially leading to inappropriate tool trajectories or unintended tool--tissue interaction. Therefore, a perception system such as TransGHOST, which can recover visual context beneath the instrument in real time, is needed. We investigate whether the occluded part of the scene can be recovered in real time from previously observed pixels. Our method excludes synthesis by construction, and we measure its reconstruction fidelity.

Augmented-reality systems in other surgical fields can prevent virtual overlays from covering the tools \cite{DeBacker2023, Hofman2024Deocclusion}, but these systems address the visibility of the instrument rather than the tissue it conceals. Existing work on general-purpose video inpainting has primarily focused on generative models, so the content reconstructed within the occluded region is synthesised rather than observed and cannot be treated as direct visual evidence during surgery.

Our objective is to restore occluded tissue in real time without synthesising anatomy and to mark pixels that lack support from an observed reference. TransGHOST's safe mode (Figure~\ref{fig:ref_inpaint_pipeline}) leaves each output value unchanged or obtains it by geometrically resampling a recent recorded frame. It also provides per-pixel provenance and abstains when the reference cannot support a reconstruction. All main fidelity, latency and provenance results refer to this output. On the new veterinary clinical corpus, its median and 95th-percentile latencies are both within the sub-100~ms window for live surgical guidance.

This study makes five contributions to reference-guided reconstruction for surgical video. (i) We show that instrument occlusion in this setting can be resolved without synthesising tissue: each restored value is resampled from a recorded frame under transforms fitted only to visible regions, and it carries the reference frame and source coordinate it came from, so a downstream component can separate observed pixels from unsupported ones. (ii) We define a region-aware reference-selection policy that chooses, from a bounded causal buffer, the newest frame whose instrument mask leaves the restoration region visible, and we compare its masked PSNR with that of selecting the most recent frame. (iii) We construct a benchmark in which candidate references may themselves be occluded. (iv) We establish an evaluation protocol comprising scoring within the masked region, separation of each scoring target from its reference, reference-free controls, a cutout-disjoint held-out cohort and an analysis of fidelity against reference age. (v) We introduce a pixel-annotated ophthalmic corpus labelled for both instruments and anatomy, enabling the controlled semi-synthetic evaluation required for reference-guided reconstruction (Section~\ref{ssec:corpus}).

The occluded-reference benchmark (iii) is necessary to evaluate the selection policy (ii) because conventional composited benchmarks typically supply instrument-free candidate frames. When every candidate leaves the target region visible, a region-aware policy becomes indistinguishable from recency.

Scored over the occluded region alone, the system reaches \vSelectedPSNR~dB masked PSNR and \vSelectedSSIM{} masked SSIM at a 95th-percentile latency of \vSelectedPninetyfive~ms. On the held-out cohort using training-excluded instrument cutouts, it reaches \vHeldPSNR~dB masked PSNR with a 95th-percentile latency of \vHeldPninetyfive~ms.

\section{Related Work}
\label{sec:related}

\subsection{Robot-assisted intraocular surgery and perception under occlusion}
\label{ssec:related_robotics}

Robotic assistance for intraocular microsurgery increasingly combines manipulation with intraoperative imaging \cite{Iordachita2022}. Robot-assisted intraocular surgery has been performed in patients \cite{Edwards2018}, virtual fixtures for a handheld micromanipulator have been generated in real time from microscope video \cite{Becker2013}, and a policy learned from expert demonstrations has navigated a tool to retinal targets in an eye phantom \cite{Kim2020}. In cataract surgery, semi-automated lens extraction guided by optical coherence tomography has been demonstrated in post-mortem porcine eyes \cite{Chen2018IRISS}. In such systems the controlled instrument also occludes part of the image used to guide it, so the tissue on which it acts is hidden from the components that track, register or constrain it.

Perception methods for robot-assisted surgery already contend with tool occlusion within their own models. Real-time simultaneous localisation and mapping of the retinal vasculature for robot-assisted intraocular microsurgery has been demonstrated under conditions that include tool occlusion \cite{Braun2018EyeSLAM}. In robotic minimally invasive surgery, dynamic reconstruction from stereo endoscopic video has used tool segmentation and temporal aggregation to recover the scene obstructed by the tool \cite{Long2021EDSSR}, and neural rendering of deformable tissue has used tool-mask-guided ray casting to handle the occlusion \cite{Wang2022EndoNeRF}. TransGHOST addresses the related problem of tool-occluded scene perception at the image level, from a single camera view and without depth estimation or a scene model.

\subsection{Ophthalmic datasets and real-time segmentation}

Widely used public ophthalmic datasets primarily annotate anatomical structures rather than surgical instruments. DRIVE, STARE and CHASE\_DB1 are established benchmarks for retinal vessel and structure segmentation, but none labels forceps, cutters or lasers \cite{Kovacs2021, Hatamizadeh2022}. Studies that require instrument labels often rely on proprietary collections, including datasets for cataract procedures \cite{Morita2020} and vitrectomy \cite{Nespolo2022}. Instrument localisation in eye surgery has also been addressed with single-shot detectors \cite{Sogabe2022Detection}, but a bounding box does not delimit the occluded pixels that reference-guided restoration has to replace.

Several groups have adapted lightweight convolutional networks to the latency requirements of the operating room. PhacoTrainer tracks cataract instruments at 30~fps \cite{Yeh2023PhacoTrainer}, and YOLACT++ jointly segments the retina and instruments at approximately 87~fps \cite{Nespolo2022}. Compressed U-Nets delineate optical coherence tomography layers with sub-4~ms inference \cite{Borkovkina2020}, and PseudoSegRT operates below 1~ms per frame for semi-supervised optical coherence tomography guidance \cite{Huang2023}. All of these methods address real-time segmentation rather than restoration of the tissue obscured by the instruments.

\subsection{Video inpainting}

Flow-guided methods propagate motion edges with optical flow to preserve temporal boundaries, as in Flow-edge Guided Video Completion \cite{fgvc20}. The end-to-end E2FGVI model \cite{e2fgvi22} instead learns flow completion and feature propagation jointly, improving PSNR and SSIM at per-frame runtimes of about 0.12~s at 240p. The transformer-based ProPainter \cite{zhou2023pro} combines recurrent flow completion with sparse attention to improve long-range temporal coherence, and its code and weights are public. All three methods are generative and require future frames, so their reconstructed regions may depict anatomy that was never observed and they cannot operate online during surgery.

\subsection{Reference-guided de-occlusion in surgery}

Reference-guided inpainting supplies an image in which the occluded region is visible, thereby reducing reliance on generative completion. Koreeda et~al.\ \cite{Koreeda2016} rendered instruments virtually transparent at 62.4~ms per frame by warping a second endoscope's view into the hidden region. This construction does not synthesise image content but requires a second endoscope. Hasan et~al.\ subsequently applied reference-guided tool removal to laparoscopy, initially using a pool of frames captured before the instrument entered \cite{Hasan2021}. Their later system selects, for each frame, the candidate whose visible pixels most closely match the current view near the instrument. A companion method in the same work instead accumulates a background map over time \cite{Hasan2023}. General-purpose video inpainting systems also use references, but refresh them at a fixed frame stride rather than according to whether the target region is visible \cite{Li2024optimised}.

Xie et~al.\ \cite{Xie2023} address instrument occlusion in minimally invasive endoscopic surgery. Their method is designed for large, sustained tool-channel occlusions and blends a generative reconstruction of the current frame with information from previously observed occlusion-free frames. This hybrid design can return a complete image when the stored background is insufficient. However, generated content may enter the final blend, and the design aims at a complete image rather than at separating observed from inferred pixels. TransGHOST addresses a different regime of smaller, shorter and repeated occlusions in ophthalmic microsurgery. Its safe mode resamples recorded image content, provides a per-pixel provenance map and marks unsupported pixels as abstentions. On the occluded-reference benchmark, its region-aware selector improves masked PSNR by \vOccGain~dB over recency and \vOccGainOverPublished~dB over the published selection criterion evaluated here.

Xie et~al.\ validate their method on clinical endoscopic video and report a mean throughput of 12.02 frames per second. They identify real-time streaming as future work, and their masked metrics characterise the single-frame restoration component during training rather than the complete hybrid output. Our evaluation measures the full TransGHOST pipeline with synchronised per-frame latency percentiles, verifies that the scoring target differs from the supplied reference, and includes a pre-declared cohort using instrument cutouts excluded from segmentation training. Because the datasets and protocols differ, the results of the two studies cannot be compared directly.

\subsection{Occlusion handling for surgical perception and guidance}

In other surgical fields, augmented-reality systems handle occlusion by controlling where overlays are drawn. De Backer et~al.\ \cite{DeBacker2023} delineate robotic instruments in renal surgery so that augmented-reality meshes do not cover them, and Celdr\'an et~al.\ \cite{Celdran2025} review comparable assistance in laparoscopy. These systems preserve the visibility of the instrument rather than revealing the tissue behind it. TransGHOST instead reconstructs that tissue within the 100~ms limit reported as acceptable for surgical video feedback in telesurgery and laparoscopy \cite{Nankaku2022, Kumcu2017}.

Real-time occlusion handling has also been demonstrated in ophthalmology. Liu et~al.\ \cite{Liu2025DALK} describe a tracking and inpainting network for augmented-reality guidance in deep anterior lamellar keratoplasty, reporting 16.26~ms for inpainting inference and 35.14~ms end to end. Their inpainting network is generative. TransGHOST targets traceable reconstruction: its safe mode resamples observed image content and requires only past frames. These studies establish several capabilities relevant to surgical de-occlusion, but the intended intraoperative setting imposes a specific set of operational requirements on the complete reconstruction pipeline. It must operate causally from a single camera stream, reconstruct occluded tissue without introducing unsupported synthetic content, identify when a reference cannot support a reconstruction, retain the source of each restored pixel, and operate within latency compatible with live surgical use. TransGHOST is designed around these requirements. It detects the occluding instrument, selects a causally available reference in which the target region is visible (or, if none is fully visible, the least occluded one) \cite{Conze2016, Li2020optical}, aligns it to the current frame, reconstructs supported pixels only from recorded image content, and outputs confidence, provenance, and abstention together with the restored image. To our knowledge, fidelity as the reference ages and preservation outside the restoration mask remain underexplored in surgical studies.

\section{Materials and Data}
\label{sec:data}

\begin{figure}[pos=t]
\centering
\includegraphics{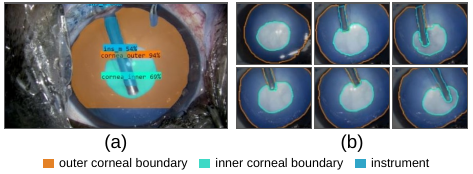}
\caption{Dataset preparation: (a) an annotated surgical frame with labelled corneal boundaries and instruments, and (b) further annotated frames from the training corpus.}
\label{fig:seg-annotate}
\end{figure}

\subsection{Surgical video corpus}
\label{ssec:corpus}

Reference-guided restoration requires labels for both the visible anatomy and the instruments that occlude it. Existing cataract datasets provide part of this information: \textsc{CaDIS} \cite{CaDIS} labels corneal anatomy and a fine-grained instrument taxonomy, whereas \textsc{Cata7} \cite{Ni2019Cata7} labels instruments. They were designed for segmentation research rather than for evaluating restoration, which needs paired clean targets and temporally separated references. Our study also requires material from which controlled semi-synthetic composites can be constructed. We therefore built a veterinary clinical corpus with two groups of labels: the inner and outer corneal boundaries and all surgical instruments present.

The corpus comprises 13 high-definition ($1920\times1080$, 29.97\,fps) recordings of canine cataract surgery. Recordings are assigned to roles before any frame is used: six supply the training composites, one the validation set and three the held-out cohort (Section~\ref{ssec:heldout}), two are sealed for future evaluation and one is unassigned. This partition governs every TransGHOST result. The two learned models trained in this work, the TransRef-inspired baseline (Section~\ref{ssec:inpaint_survey}) and the refinement in Appendix~\ref{app:quality}, were instead trained on the triplets of Section~\ref{ssec:triplets}, which draw on eleven recordings, including the validation recording and two of the held-out recordings. Their scores on these recordings may therefore be optimistic. Each recording covers a single operative step. The ten recordings behind the results reported here cover nucleus sculpting, division and rotation together with anterior-chamber maintenance, and the instruments entering the field across them are a phacoemulsification tip, a hook and a 27-gauge blunt cannula. One recording in the held-out cohort contains a tip and a hook at the same time. Each recording was subsampled to retain clinically salient moments while reducing frame redundancy.

\subsection{Ethics and approvals}
\label{ssec:ethics}

Clinical surgical images were obtained from canine patients undergoing phacoemulsification for the treatment of cataracts. The images were acquired using a surgical microscope (Leica M844 F40, Leica Microsystems) following pharmacological pupil dilation. The original surgical videos were recorded at a resolution of $1920\times1080$ pixels and a frame rate of 29.97\,fps.

All surgical procedures were performed as part of routine veterinary care, and no additional invasive interventions were conducted for research purposes. The use of clinical image data for research was reviewed and approved by the Animal Clinical Research Ethics Review Committee of the Faculty of Agriculture, Tottori University (Approval No.\ \mbox{R8-006}).

To protect the confidentiality of veterinary clinical records and minimize the use of personal clinical information, the research dataset was limited to surgical image data required for this study. Accordingly, demographic information, including breed, sex, and age, was not included in the dataset provided for research.

\subsection{Imaging field and clinical scope}
\label{ssec:scope}

The recordings show cataract surgery on the operative eye from a single high-magnification camera. The camera provides a frontal view in which the eye occupies most of the frame, with periocular tissue and the surgical drape visible at the margins and instruments entering laterally. The two anatomy classes delineate the inner and outer corneal boundaries, within which TransGHOST reconstructs tissue occluded by an instrument.

Safe-mode reconstruction does not depend on identifying the tissue hidden by the instrument. Each restored value is resampled from an earlier frame recorded by the same camera, and the geometric transforms are fitted exclusively to regions visible in the current frame.

\subsection{Annotation}
\label{ssec:annotation}

Annotation proceeded in three stages. Figure~\ref{fig:seg-annotate} shows examples of the labelled frames.

\emph{(i) Manual seeding.} We sampled a seed set of \(\tilde{N}=400\) frames from the development videos and manually annotated each frame with polygons.

\emph{(ii) Manual expansion.} We manually expanded the seed to \(1{,}760\) polygon-annotated frames, the set on which the segmentation architectures of Section~\ref{ssec:seglogic} are compared in Section~\ref{ssec:seg_results}. All counts of ``hand-labelled frames'' in this paper refer to this set rather than to the 400-frame seed.

\emph{(iii) Automatic expansion and compositing.} We enlarge the hand-labelled set through confidence-gated pseudo-labelling and semi-synthetic compositing (Section~\ref{ssec:expansion}). Temporal jitter of \(\pm1\) frame adds further variation. Table~\ref{tab:dataset_stats} gives the resulting totals. Source videos are assigned to data splits before expansion, and every derived frame remains in the same split as its source.

\paragraph{Class taxonomy.}
The annotation defines six classes, divided into anatomy and tools:
\[
\mathcal{C}=\{\text{\textbf{cor}}_{\text{outer}},\,
              \text{\textbf{cor}}_{\text{inner}},\,
              \text{\textbf{ins}}_m,\,
              \text{\textbf{ins}}_s,\,
              \text{\textbf{ins}}_t,\,
              \text{\textbf{ins}}_{\text{other/reflection}}\},
\]
where \textbf{cor} denotes the corneal boundaries that delimit the intra-ocular field of view and \textbf{ins} denotes the surgical instruments. Specular reflections are predominantly instrument-associated in these recordings and are therefore grouped with unclassified instruments for restoration.

Class frequencies vary widely across this annotated set. The two corneal classes appear in $1{,}725$ frames each and the main instrument class $\text{ins}_m$ in $1{,}259$, while $\text{ins}_s$, $\text{ins}_t$ and $\text{ins}_{\text{other/reflection}}$ appear in $250$, $135$ and $23$ frames, respectively. Because restoration requires the location of instrument pixels rather than instrument identity, the four tool classes are merged into a single instrument channel for every restoration experiment. This merged channel pools all annotated tool pixels during training, while the annotation files retain the per-class taxonomy for other applications.

\subsection{Dataset expansion pipeline}
\label{ssec:expansion}

\paragraph{Automatic labelling with the trained segmenter.}
Given an unlabelled frame \(x\), predictions from the trained segmentation model \(f_\theta\) are retained at a confidence threshold of \(\tau=0.9\) and converted to pseudo-label masks. This process adds \(N_{\text{pseudo}}\approx20\text{k}\) pseudo-labelled images to the training pool.

\paragraph{Semi-synthetic compositing.}
For each accepted mask \(\mathcal{M}_j\), we select a background frame \(\mathcal{B}_k\) whose own instrument mask does not overlap it, \(\mathcal{M}_j\!\cap\!\mathcal{M}_{\!B_k}=\varnothing\), and generate a composite as
\[
I_{jk}^{\text{syn}}
  = \mathcal{B}_k\odot(1-\mathcal{M}_j)\;+\;I_j\odot\mathcal{M}_j .
\]
Each composite receives one randomly selected augmentation: a flip, elastic deformation, colour jitter, a rotation with \(\theta\sim\mathcal{U}[-20^\circ,20^\circ]\), or a scaling with \(s\sim\mathcal{U}[0.8,1.2]\). The process yields \(40\text{k}\) synthetic training images and \(6\text{k}\) synthetic test images. Together with the real frames, these produce the totals in Table~\ref{tab:dataset_stats}.

\begin{table}[pos=t]
\centering\small
\caption{Corpus size after each expansion stage. The real column counts recorded frames carrying hand or confidence-gated pseudo labels and the synthetic column counts semi-synthetic composites. Source videos are assigned to a split before expansion, so no derived frame crosses a split boundary. This is the segmentation partition and is separate from the restoration evaluation sets of Section~\ref{ssec:protocol_summary}.}
\label{tab:dataset_stats}
\begin{tabular}{cccc}
\toprule
Split & Real & Synthetic & Total images\\
\midrule
Train & 20,000 & 40,000 & 60,000\\
Test  &  4,000 &  6,000 & 10,000\\
\bottomrule
\end{tabular}
\end{table}

\begin{figure}[pos=t]
\centering
\includegraphics{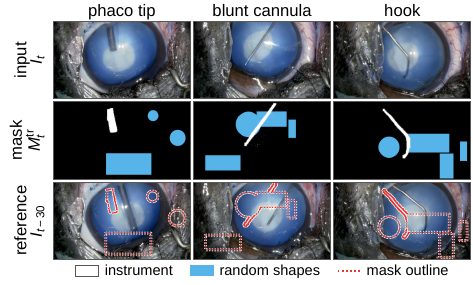}
\caption{Training triplets from three development videos, one per instrument. Each removal mask $M^{\mathrm{tr}}_t$ unites the instrument mask (white) with one to four random rectangles or circles (blue), which may overlap one another and the instrument. The mask is binary, and the two colours only distinguish its parts. The reference $I_{t-30}$ is the frame 30 frames earlier, and the dotted outline marks the mask region in it. In these examples the instrument has moved away from its own masked region.}
\label{fig:data-generate-ref}
\end{figure}

\subsection{Training triplets}
\label{ssec:triplets}

For training the learned components, we build the triplets shown in Figure~\ref{fig:data-generate-ref}, \(\langle I_t,\,M^{\mathrm{tr}}_t,\,I_{t-\Delta_{\mathrm{tr}}}\rangle\). The offset is \(\Delta_{\mathrm{tr}}=30\) frames for triplets taken from recorded video and \(22\) frames for triplets in which recorded instrument motion is overlaid on a static background frame. As at deployment, the reference precedes the frame it restores, and the offset introduces temporal variation between the target and its reference. The training mask \(M^{\mathrm{tr}}_t\) is the instrument mask united with one to four randomly placed, axis-aligned rectangles or circles. Rectangle sides are drawn between 5\% and 33\% of the frame width and height, and circle radii between 2.5\% and 16.7\% of the shorter side. The added shapes encourage the learned components to restore varied regions rather than a single silhouette. These triplets train the optional learned variant in Appendix~\ref{app:quality} and the baseline generator of Section~\ref{ssec:inpaint_survey}. The deployed safe-mode fill contains no network and requires no training data (Section~\ref{ssec:safe}).

At inference, the bounded causal buffer of Section~\ref{ssec:refsel} retains recent frames rather than sampling at a fixed stride. The regional overlap rule selects among them and can bypass a sequence of references in which the target region remains occluded.

\section{The TransGHOST Method}
\label{sec:method}

\subsection{Problem formulation and pipeline overview}
\label{ssec:formulation}

TransGHOST recovers occluded anatomy through a three-stage pipeline applied to each frame \(I_t\) and summarised in Algorithm~\ref{alg:transghost}. A lightweight segmentation model produces an instance mask of all surgical tools. A region-aware reference-selection policy scans a bounded causal buffer for a recent frame in which the region now hidden was visible. A fill stage then reconstructs that region from the selected reference. All weights are fixed during inference, and the equations and timings describe inference only.

TransGHOST reconstructs the occluded region through safe mode (Section~\ref{ssec:safe}), which geometrically registers an earlier reference to the current frame and resamples it within the occluded region. Every transform is estimated outside the current instrument mask, where the current frame shows tissue rather than the instrument, and no restored pixel is drawn from a coordinate hidden in the reference. Bilinear resampling makes each restored value a weighted average of neighbouring intensities in one recorded frame, so safe mode is non-generative by construction. Safe mode is the default configuration and produces the output evaluated throughout the main paper. Appendix~\ref{app:quality} evaluates an optional learned refinement as a separate design variant.

\begin{algorithm*}[t]
\caption{\textsc{TransGHOST} processing for one frame: instrument segmentation, regional reference selection and gated geometric compositing.}
\label{alg:transghost}
\textbf{Input}: video stream $\{I_t\}_{t=1}^{T}$\\
\textbf{Parameters}: segmentation confidence $\tau$, allowed classes $\mathcal{C}$, compositing confidence threshold $\tau_c$, buffer length $L$\\
\textbf{Output}: restored stream $\{\hat I_t\}_{t=1}^{T}$, with per-pixel confidence $c_t$, abstention mask, and provenance map
\begin{algorithmic}[1]
\STATE Initialise causal buffer $\mathcal{H}\gets\emptyset$
\FOR{$t \gets 1$ \textbf{to} $T$}
    \STATE $M_t \gets \textproc{InstanceSegment}\!\left(I_t;\,\tau,\,\mathcal{C}\right)$ \COMMENT{instrument mask, Section~\ref{ssec:mask}}
    \IF{$M_t = \emptyset$ \textbf{or} $\mathcal{H} = \emptyset$}
        \STATE $\hat I_t \gets I_t$ \COMMENT{nothing to restore, or no past frame yet}
    \ELSE
        \STATE $(I_{\mathrm{ref}},M_{\mathrm{ref}}) \gets \operatorname*{arg\,min}_{(I',M')\in\mathcal{H}} \bigl(\lVert M_t\wedge M'\rVert_0 / \lVert M_t\rVert_0,\ -t'\bigr)$ \COMMENT{Eq.~(\ref{eq:disjoint})}
        \STATE $(\hat I_t,\,c_t) \gets \textproc{SafeComposite}\!\left(I_t,\,M_t,\,I_{\mathrm{ref}},\,M_{\mathrm{ref}}\right)$ \COMMENT{Section~\ref{ssec:safe}}
        \STATE Retain $I_t(p)$ wherever $c_t(p) < \tau_c$ \COMMENT{abstention, Eq.~(\ref{eq:blend})}
    \ENDIF
    \STATE Emit $\hat I_t$ with $c_t$, the abstention mask and the provenance map
    \STATE Append $(I_t,\,M_t)$ to $\mathcal{H}$, keeping the most recent $L$ entries
\ENDFOR
\end{algorithmic}
\end{algorithm*}

\subsection{Instrument segmentation}
\label{ssec:mask}

The restoration stage requires a binary occluder mask rather than the identity of each instrument. The deployed segmentation model therefore collapses the four annotated tool classes $\{\text{ins}_{m},\text{ins}_{s},\text{ins}_{t},\text{ins}_{\text{other/reflection}}\}$ into a single \emph{instrument} channel during training, so that the examples of all four classes are pooled (Section~\ref{sec:data}).

Masks smaller than five pixels are discarded before the remaining masks are merged and dilated:
\[
M_t^{\mathrm{raw}}=\bigvee_k m_{t,k},\qquad
M_t=\operatorname{dilate}(M_t^{\mathrm{raw}},\mathbf{S}_{14}),
\]
with $\mathbf{S}_{14}$ a $14{\times}14$ pixel kernel that provides a restoration margin around detected tool boundaries. The final dilated mask is used throughout end-to-end evaluation, so the reported precision, recall, IoU and false-positive MAE characterise the complete mask passed to the compositor. Segmentation takes a median of \vSegMedianMs~ms per $1920{\times}1080$ frame and \vSegPninetyfiveMs~ms at the 95th percentile. These values were measured during the deployed timing evaluation on the same RTX 5000 Ada used for every other timing in this paper.

\subsection{Reference selection}
\label{ssec:refsel}

\begin{figure}[pos=t]
\centering
\includegraphics{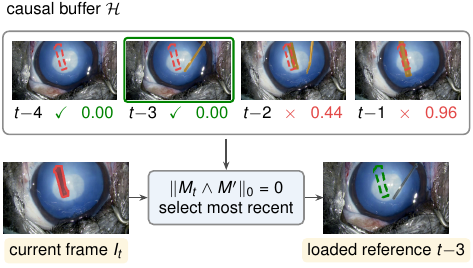}
\caption{Regional reference selection over the causal buffer. The overlap rule selects the newest candidate whose instrument mask leaves the target region visible.}
\label{fig:buffer}
\end{figure}

A bounded causal buffer $\mathcal{H}=\{(I_{t-i},M_{t-i})\}$ stores the $L=12$ most recent frames and their predicted instrument masks (Figure~\ref{fig:buffer}). Only frames with indices earlier than $t$ enter the buffer, preventing the selector from using future information, and the fixed length bounds both memory and the per-frame selection cost independently of sequence duration. At 29.97\,fps the buffer spans $0.4$~seconds, which is the reference-age interval characterised in Section~\ref{ssec:lag}.

Candidate references are ranked by the overlap between their instrument masks and the current restoration region in image coordinates:
\begin{equation}
(I_{\mathrm{ref}},M_{\mathrm{ref}})=
\operatorname*{arg\,min}_{(I',M')\in\mathcal{H}}
\Bigl(\tfrac{\lVert M_t\wedge M'\rVert_0}{\lVert M_t\rVert_0},\ -t'\Bigr).
\label{eq:disjoint}
\end{equation}
The two terms are ordered lexicographically. The selector first chooses the most recent reference with zero overlap. If every reference covers part of the region, it instead chooses the one with the smallest overlap. Equation~(\ref{eq:disjoint}) is a low-cost visibility score, computed before registration, for the short causal intervals considered here. Final pixel support is determined after registration: Eq.~(\ref{eq:blend}) checks whether every sampled source coordinate lies within the reference image and outside $M_{\mathrm{ref}}$. The selector therefore only ranks candidates, and the confidence, abstention and provenance outputs govern which pixels enter the output, even when the two criteria identify different supports. This regional criterion improves masked PSNR by \vOccOverlapOverRanked~dB over the composite frame-quality score evaluated in Section~\ref{ssec:selection}. It also requires only one mask intersection per reference, whereas an appearance-based comparison evaluates image values over a region.

\paragraph{Reference-selection behaviour across evaluation sets.}
The reported protocol uses Eq.~(\ref{eq:disjoint}). On the validation and held-out sets every supplied reference is instrument-free, so the overlap term carries no information about the target region and the rule follows its recency tie-break on essentially every sample. The held-out metrics are identical under the two rules, and on unobstructed references the two rules select the same reference on \vSigCleanIdentical{} of the \vSigN{} validation samples. The $0.05$~dB separation reported in Section~\ref{ssec:selection} comes from the remaining five, with recency marginally ahead. On the occluded-reference benchmark of Section~\ref{ssec:selection}, where candidates can hide the region, the overlap rule improves masked PSNR by \vOccGain~dB over recency.

If the selected reference still covers part of $M_t$, the system reconstructs the supported pixels and retains the current-frame values elsewhere. The abstention mask records the remaining region explicitly, preventing pixels from one instrument from being transferred into the location of another. Section~\ref{ssec:selection} quantifies this behaviour when the candidate pool provides partial rather than complete support.

\subsection{Safe-mode reconstruction}
\label{ssec:safe}

Safe mode reconstructs $M_t$ by locating the corresponding image content in $I_{\mathrm{ref}}$ and resampling it. Within $M_t$, the current frame shows the instrument rather than the underlying tissue, so measurements from this region would track instrument motion and are excluded from every fit.

Registration proceeds through three stages, from global alignment to local correction, and the two masks are used at different stages. The global fit is constrained by $M_t$ alone, because a whole-frame criterion is dominated by the large unobstructed area and remains well conditioned when part of the reference is occluded. The reference mask $M_{\mathrm{ref}}$ then enters the two stages that read individual reference samples: it restricts the band in Eq.~(\ref{eq:residual}) and the blockwise fit to coordinates that the reference can supply. The reference mask also sets the confidence to zero in Eq.~(\ref{eq:blend}) wherever the sampled source coordinate falls inside it, so that a restored pixel is drawn only from a location the reference records as tissue.

\paragraph{Global registration.}
An enhanced correlation coefficient criterion fits a $2\times3$ affine map $A_t$ from the current frame to the reference. The fit is evaluated on a decimated image, with $M_t$ excluded from the objective:
\begin{equation}
A_t=\arg\max_{A}\ \rho\!\left(I_t,\ I_{\mathrm{ref}}\!\circ\!A;\ \neg M_t\right),
\label{eq:ecc}
\end{equation}
where $\rho$ is the correlation coefficient over the valid support. Equation~(\ref{eq:ecc}) accounts for camera drift and the dominant component of tissue motion.

\paragraph{Residual translation.}
The affine fit typically leaves a residual displacement of a few pixels. We estimate this displacement by evaluating integer and then fractional translations $d$ on a narrow band $B_t$ of visible pixels outside the mask:
\begin{equation}
d_t=\arg\min_{d}\ \frac{1}{|B_t(d)|}\sum_{p\in B_t(d)}
\bigl|\,I_t(p)-I_{\mathrm{ref}}(A_t p+d)\,\bigr|.
\label{eq:residual}
\end{equation}
The valid band at offset $d$ is
\[
B_t(d)=\{p\in B_t:\ A_tp+d\ \text{is in view and outside } M_{\mathrm{ref}}\}.
\]
It contains the pixels for which the reference provides a valid sample. Normalising by $|B_t(d)|$ prevents an offset from receiving a lower error merely because it moves samples outside the image. Since $B_t$ lies outside $M_t$, the hidden region does not contribute to Eq.~(\ref{eq:residual}).

\paragraph{Per-block local offsets.}
Local tissue deformation between frames is represented by fitting an independent translation for each $N\!\times\!N$ block using the same visible-region criterion. Blocks meeting the support criterion contribute offsets that are interpolated into a continuous field $\delta_t$ to avoid seams. The comparison with dense optical flow in Appendix~\ref{app:rejected} supports estimation from visible regions outside the instrument mask.

\paragraph{Compositing.}
The reference is sampled once, at the final coordinates and at native resolution. Writing $s_t(p)=A_t p+d_t+\delta_t(p)$ for the source coordinate of pixel $p$, the sampled value is $\tilde I_t(p)=I_{\mathrm{ref}}(s_t(p))$. A per-pixel confidence $c_t$ combines registration agreement with two validity conditions, the source coordinate lying within the reference image domain $\Omega$ and outside the reference mask:
\[
c_t(p)=\min\!\left(\rho_t,1\right)^{+}\cdot
\mathbf{1}\!\left[s_t(p)\in\Omega\right]\cdot
\mathbf{1}\!\left[s_t(p)\notin M_{\mathrm{ref}}\right],
\]
where $\rho_t$ is the correlation coefficient returned by Eq.~(\ref{eq:ecc}), clamped to $[0,1]$ and set to zero if the fit does not converge. Pixels below the threshold $\tau_c$ form the abstention set and remain unmodified. A source coordinate outside the frame or inside the reference mask therefore has zero confidence, irrespective of photometric agreement. Within a frame, support is decided by geometry, and the photometric term $\rho_t$ acts as a gate on the whole frame.

For reproducibility, global registration uses a \vCfgRegScale{} decimation, \vCfgRegIterations{} iterations and \vCfgRoiPad~pixels of context around the mask bounding box. Residual translation is evaluated over a \vCfgResidualBand-pixel band, within a radius of \vCfgResidualRadius{} and at \vCfgSubpixelSteps{} sub-pixel steps. Local offsets use $N=\vCfgBlock$ pixel blocks, a search radius of \vCfgBlockRadius{} and \vCfgBlockScale{} of native resolution. The remaining thresholds are $\tau_c=\vCfgSupport$, mask-core confidence \vCfgCore{}, uncertain-mask confidence \vCfgUncertain{} and segmentation confidence \vCfgSegConfidence. The output is
\begin{equation}
\hat I_t(p)=
\begin{cases}
\tilde I_t(p) & p\in M_t \ \wedge\ c_t(p)\ge\tau_c,\\
I_t(p) & \text{otherwise}.
\end{cases}
\label{eq:blend}
\end{equation}
Thus, every pixel outside $M_t$ retains its input value. Alongside $\hat I_t$, the pipeline returns $c_t$, the abstention mask and a provenance map containing the reference frame and source coordinate for each replaced pixel.

\subsection{Latency}
\label{ssec:latency}

Under the synchronised protocol of Appendix~\ref{app:protocol}, safe mode has a median end-to-end latency of \vSelectedPfifty~ms and a 95th percentile of \vSelectedPninetyfive~ms on an RTX 5000 Ada. Both are below the 100~ms delay reported as acceptable for surgical video feedback in telesurgery and laparoscopy \cite{Nankaku2022, Kumcu2017}. Segmentation accounts for \vSegMedianMs~ms at the median and the three-stage alignment for \vAlignMedianMs~ms. The remaining time covers native-resolution compositing, confidence and provenance maps, and frame input and output. Stage medians do not sum exactly to the end-to-end median because they occur on different frames. Section~\ref{ssec:realtime} examines the tail of the latency distribution.

\section{Experimental Design}
\label{sec:experimental}

\subsection{Baselines}
\label{ssec:inpaint_survey}

The baseline methods differ in the context available to them and in whether they synthesise unobserved content. \emph{Single-image} models such as LaMa \cite{suvorov2021resolution} receive only the current frame and must infer the vessels and reflections behind an instrument, and their output can vary from frame to frame. The \emph{video} methods considered here propagate information through flow or attention \cite{Ouyang2021Internal,zhou2023pro}. They improve temporal coherence but use future frames and remain generative when an occlusion persists. \emph{Reference-guided} methods instead receive an earlier frame in which the hidden region is visible, allowing observed content to inform the reconstruction.

LaMa is included as the reference-free baseline and is evaluated under the same masked-region protocol. The real-time comparison is restricted to causal methods, whereas video inpainters that require future frames fall outside this operating condition.

\begin{figure*}[pos=t]
\centering
\includegraphics{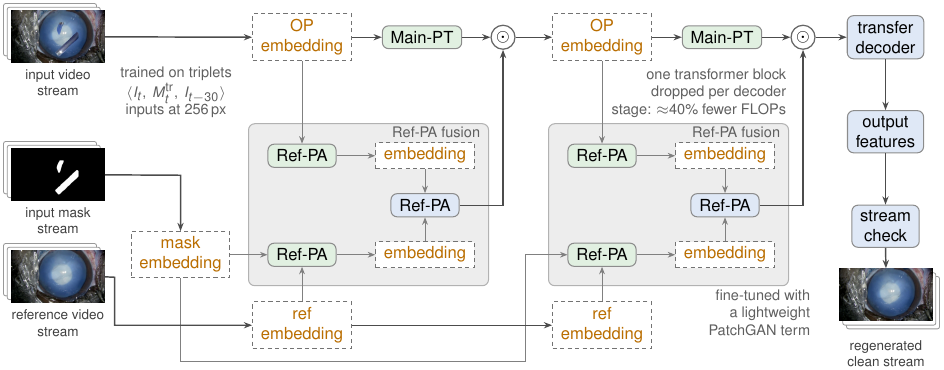}
\caption{Real-time-constrained TransRef-inspired baseline. Reference features enter through Ref-PA modules and a transfer decoder regenerates the masked stream at $256$~px.}
\label{fig:transref-arch}
\end{figure*}

We evaluate both learned and geometric forms of reference guidance. The learned comparison is a real-time-constrained adaptation of the dual-encoder reference transformer of TransRef \cite{TransRef, RGTSI}: one transformer block is removed from each decoder stage, reducing floating-point operations by about $40$~per cent, and a lightweight PatchGAN term is added during fine-tuning on the triplets of Section~\ref{ssec:triplets} (Figure~\ref{fig:transref-arch}). It operates at $256$~px. Section~\ref{ssec:lama} evaluates it using the same samples, masks, references and masked metrics as the other restoration methods.

The geometric alternative deployed by TransGHOST registers the reference to the current frame and transfers recorded image content without using a network for reconstruction. A learned fill can produce content absent from the reference and must therefore be assessed empirically, whereas geometric transfer resamples only recorded content. Section~\ref{sec:results} compares classical diffusion, reference-free learned inpainting, reference-guided learned inpainting and geometric transfer.

\subsection{Metrics}
\label{ssec:metrics}

Restoration quality is measured over the occluded region using four metrics. \textbf{PSNR} (peak signal-to-noise ratio, $\uparrow$) measures pixel fidelity and penalises large errors quadratically. \textbf{SSIM} (structural similarity index, $\uparrow$) measures luminance, contrast and structural agreement. \textbf{MAE} (mean absolute error, $\downarrow$) is reported in 0-255 intensity levels, while \textbf{LPIPS} (learned perceptual image patch similarity, $\downarrow$) measures perceptual differences in texture and structure. Where two systems are compared, they are scored on the same frames, masks and references, so the per-sample values are paired and are compared with a two-sided Wilcoxon signed-rank test rather than by their means alone.

\subsection{Segmentation-model training}
\label{ssec:seglogic}

We fine-tuned three architectures on the \(1{,}760\)-frame hand-labelled set: Detectron2 Mask R-CNN, YOLOv8 and YOLOv11. Optimiser settings were shared where the architectures permitted, with the same initial learning rate, momentum and weight decay. The training budgets differ: YOLOv8 was fine-tuned for 50 epochs at batch size 24 and YOLOv11 for 100 epochs at batch size 20, so the comparison is between the fine-tuned models rather than between architectures at a matched budget.

\begin{enumerate}
\item Mask R-CNN (Detectron2):\;
the objective is
\(\mathcal{L}= \lambda_{\text{cls}}\mathcal{L}_{\text{CE}}
             + \lambda_{\text{box}}\mathcal{L}_{\text{smooth-L1}}
             + \lambda_{\text{mask}}\mathcal{L}_{\text{BCE}}\),
with ROIAlign pooling at two resolutions (box 7\(\times\)7 and mask 14\(\times\)14). The $\lambda$ subscripts identify the branch on which each loss acts: classification (\textit{cls}), bounding-box regression (\textit{box}) and mask prediction (\textit{mask}).

\item YOLOv8 and YOLOv11:\;
both use the unified loss
\(\mathcal{L}= \mathcal{L}_{\text{obj}}+\mathcal{L}_{\text{cls}}
             + \lambda_{\text{seg}}\mathcal{L}_{\text{mask}}\).
Both were fine-tuned from the released segmentation checkpoints under the framework's automatic optimiser selection, with an initial learning rate \(lr_0=0.01\), momentum 0.937, weight decay \(5\times10^{-4}\) and an IoU assignment threshold of 0.7.
\end{enumerate}

Mask R-CNN under Detectron2 was not carried forward: its per-frame inference latency was the highest of the three and its segmentation average precision was the lowest under its own COCO evaluation protocol. Table~\ref{tab:seg_benchmark} therefore reports the two models that remained candidates, and YOLOv11 was selected. All restoration experiments use this architecture after the retraining described in Section~\ref{ssec:seg_results}, on the merged instrument channel of Section~\ref{ssec:mask}.

\begin{table}[pos=t]
\centering\small
\caption{Segmentation benchmark on the 1,760-frame hand-labelled set. Both models are scored on mask metrics at their best checkpoint, matching Figure~\ref{fig:seg-comparison}. Latency is model inference alone, measured on the same GPU and separately from the deployed segmentation stage timed in Section~\ref{ssec:realtime}.}
\label{tab:seg_benchmark}
{\setlength{\tabcolsep}{4pt}
\begin{tabular}{lccc}
\toprule
Model & F1 $\uparrow$ & mAP@.50 $\uparrow$ & Latency (ms) $\downarrow$\\
\midrule
YOLOv8-xl & 0.91 & 0.975 & 8-10 \\
YOLOv11 & \textbf{0.92} & \textbf{0.982} & \textbf{6-8} \\
\bottomrule
\end{tabular}}
\end{table}

\subsection{Evaluation protocol}
\label{ssec:protocol_summary}

Although inpainting studies often report whole-frame scores, prior surgical de-occlusion studies have used masked scoring. Xie et~al.\ \cite{Xie2023} use this approach, computing structural similarity, peak signal-to-noise ratio and root mean squared error over the filled region. Whole-frame scoring is uninformative here because the system changes only pixels within the instrument mask, which occupies a few per cent of the image. More than ninety-five per cent of such a score would come from pixels that no compared method altered, so even an inaccurate reconstruction would retain a high whole-frame PSNR. For this reason, all restoration results in Section~\ref{sec:results} use masked scoring.

A second safeguard requires the reference supplied in every semi-synthetic sequence to differ from the clean frame used as the scoring target. A pre-evaluation check excludes any sample that violates this condition. Without it, direct copying of the target would receive a perfect score and the experiment could not distinguish reconstruction from leakage. Masked scoring follows prior work, whereas the per-sample check that the target differs from its reference is specific to this protocol.

Unless stated otherwise, evaluation uses 178 native-resolution ($1920\times1080$) samples from clips excluded from training. Mask areas range from $0.1$ to $4.6$~per cent of the frame, with a median of $0.9$~per cent. Ground-truth masks isolate the restoration stage, whereas predicted masks evaluate the deployed system. This is the default validation set. Section~\ref{ssec:heldout} adds a pre-declared held-out cohort, and Appendix~\ref{app:protocol} gives the metric definitions and timing protocol.

\section{Results}
\label{sec:results}

Segmentation results use the annotated corpus from Section~\ref{sec:data}. Restoration results follow the protocol in Section~\ref{ssec:protocol_summary} and are measured at native high-definition resolution ($1920\times1080$\,px) on one GPU.

\begin{figure}[pos=t]
\centering
\includegraphics{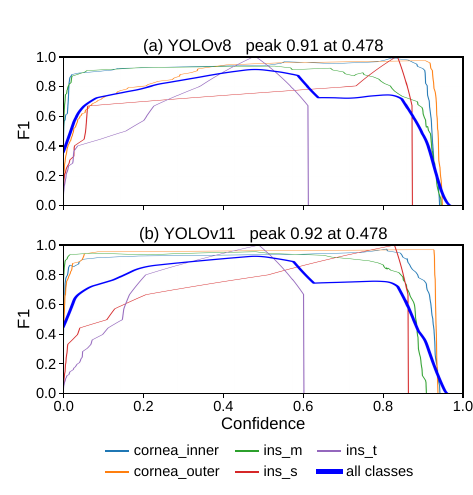}
\caption{Mask F1 against detection confidence for (a) YOLOv8 and (b) YOLOv11, from the validation run of each fine-tuned model. Thin curves are the five classes with enough annotated frames for a stable sweep and the bold curve aggregates all classes, with its maximum given in each panel title.}
\label{fig:seg-comparison}
\end{figure}

\subsection{Segmentation backbone and mask quality}
\label{ssec:seg_results}

We compared three segmentation architectures for accuracy and latency (Section~\ref{ssec:seglogic}). Figure~\ref{fig:seg-comparison} shows the mask F1 confidence sweep for YOLOv8 and YOLOv11, reproduced from the validation output of each run. On mask metrics the two models are close: F1 is 0.91 against 0.92 and mAP$_{50}$ is 0.975 against 0.982, and the training budgets behind them differ as stated in Section~\ref{ssec:seglogic}. The accuracy margin is therefore too small to decide the choice on its own. YOLOv11 was selected on inference latency, at 6-8~ms per frame against 8-10~ms, and on label stability: during visual inspection the other two architectures switched labels more often between similar instrument classes.

\begin{figure}[pos=t]
\centering
\includegraphics{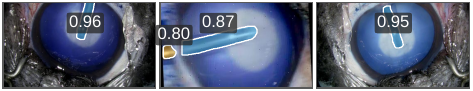}
\caption{Instrument segmentation by the deployed model on validation frames, with per-instance confidences. The model predicts the merged instrument channel of Section~\ref{ssec:mask}.}
\label{fig:seg-output}
\end{figure}

\paragraph{Segmentation model used for the restoration results.}
Table~\ref{tab:seg_benchmark} compares architectures on the multi-class annotations rather than reporting the final segmentation model. The deployed segmentation stage timed in Section~\ref{ssec:realtime} additionally assembles, merges and dilates the mask at $1920\times1080$, which accounts for its larger median. For restoration, the selected YOLOv11 backbone is retrained at a $512$~px input resolution on the merged instrument channel from Section~\ref{ssec:mask}, using the semi-synthetic augmentation in Section~\ref{ssec:expansion}. On its mixed validation set of real and semi-synthetic frames, this model reaches mask precision \vSegPrecision{}, recall \vSegRecall{} (F1 \vSegFOne{}) and mAP$_{50-95}$ \vSegMapCoarse{}. This set is distinct from the 178-sample restoration evaluation, and Figure~\ref{fig:seg-output} shows representative validation predictions.

Because restoration depends on the predicted mask, we also measure segmentation on instrument cutouts excluded from training. On the cutout-disjoint cohort in Section~\ref{ssec:heldout}, recall is \vHeldRecall{}, precision is \vHeldPrecision{} and IoU is \vHeldIoU{}. Pixel recall directly measures instrument coverage, and the held-out value corresponds to a predicted restoration mask covering \vHeldInstrumentCoverage~per cent of annotated instrument pixels. Precision and false-positive MAE separately characterise the added restoration margin beyond the annotation.

{\setlength{\tabcolsep}{3.5pt}\begin{table*}[pos=t]
\centering\small
\caption{Ablation of reference guidance and alignment on 178 validation samples. All rows use the same samples, masks and evaluation code.}
\label{tab:ablation}
\begin{tabular}{lcccccccc}
\toprule
& \multicolumn{3}{c}{Ground-truth masks} & \multicolumn{5}{c}{Predicted masks (deployed)}\\
\cmidrule(lr){2-4}\cmidrule(lr){5-9}
Fill rule or stage & PSNR $\uparrow$ & SSIM $\uparrow$ & MAE $\downarrow$ & PSNR $\uparrow$ & SSIM $\uparrow$ & MAE $\downarrow$ & Bnd.\ MAE $\downarrow$ & FP MAE $\downarrow$\\
\midrule
\multicolumn{9}{l}{\emph{Reference-free and unaligned controls evaluated with ground-truth masks}}\\
Occluded input \emph{(no restoration)} & 13.43 & 0.485 & 46.21 & \multicolumn{5}{c}{-} \\
Telea inpainting \emph{(no reference)} & 28.47 & 0.867 & 7.64 & \multicolumn{5}{c}{-} \\
Navier-Stokes inpainting \emph{(no reference)} & 28.58 & 0.867 & 7.43 & \multicolumn{5}{c}{-} \\
Unaligned reference transfer \emph{(unaligned)} & 30.26 & 0.863 & 6.73 & \multicolumn{5}{c}{-} \\
\midrule
\multicolumn{9}{l}{\emph{TransGHOST alignment components, cumulative}}\\
Global affine registration only & 33.16 & 0.905 & 4.76 & 30.44 & 0.889 & 5.29 & 2.98 & 5.42 \\
$+$ residual translation & 34.71 & 0.926 & 3.93 & 31.51 & 0.913 & 4.42 & 2.47 & 4.46 \\
$+$ sub-pixel refinement & 34.85 & 0.928 & 3.88 & 31.57 & 0.914 & 4.40 & 2.46 & 4.44 \\
$+$ per-block local offsets & \textbf{35.34} & \textbf{0.932} & \textbf{3.48} & \textbf{31.89} & \textbf{0.915} & \textbf{4.10} & \textbf{2.24} & \textbf{3.96} \\
\bottomrule
\end{tabular}
\end{table*}
}

\subsection{Contribution of the reference and alignment components}
\label{ssec:ablation}

\begin{figure*}[pos=t]
\centering
\includegraphics{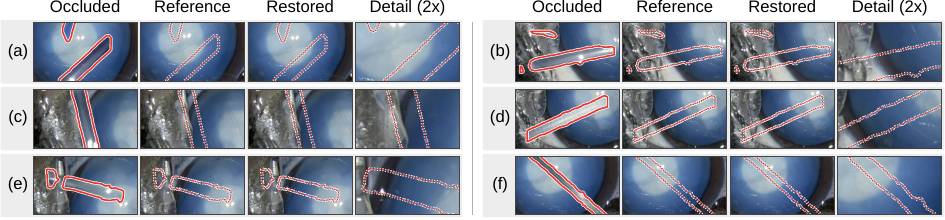}
\caption{Safe-mode results on six validation samples, labelled (a) to (f), each set on its own tinted strip of four panels. Within a strip the panels are the occluded frame, the reference selected for it, the restored composite produced by the deployed configuration, and a twofold magnification of the restored region. The solid outline traces the instrument mask recorded for that sample. The dotted outline marks that same boundary at identical image coordinates on the other three panels, where the region is unobstructed in the reference and the instrument no longer appears in the output.}
\label{fig:qualitative}
\end{figure*}

Figure~\ref{fig:qualitative} shows representative safe-mode outputs. Table~\ref{tab:ablation} tests whether reference guidance improves reconstruction and how much each alignment stage contributes. Its upper block contains reference-free controls, while the lower block adds alignment components sequentially and reports results with ground-truth masks and with predicted masks end to end. Every row uses the same samples and evaluation code. The controls are reported only with ground-truth masks because they do not include an instrument-segmentation stage.

The ablation shows that alignment enables reference guidance to improve fidelity. Telea's fast-marching method and Navier-Stokes inpainting propagate neighbouring content into the mask and achieve masked PSNR values of \vTelea~dB and \vNavierStokes~dB, respectively, on the same validation samples with ground-truth masks. Direct transfer from an unaligned reference achieves \vPaste~dB masked PSNR, approximately \vPasteGainOverInpainting~dB above Navier-Stokes, the higher-PSNR reference-free control, while registration also raises masked SSIM by compensating for tissue motion between frames. The residual-translation correction contributes the largest gain beyond the affine fit: approximately \vResidualTranslationGain~dB in masked PSNR with ground-truth masks. Sub-pixel refinement increases masked PSNR by approximately $0.1$~dB, and per-block local offsets add a further $0.5$~dB, producing \vCeiling~dB with ground-truth masks and \vSelectedPSNR~dB in the deployed configuration.

Each component improves fidelity under both mask conditions, while false-positive MAE decreases from \vAffineFPMAE{} to \vSelectedFPMAE. The approximately \vSegmentationGap~dB difference between the ground-truth-mask and deployed columns shows that mask estimation has the greatest measured influence on end-to-end fidelity. The \vCeiling~dB ground-truth-mask result also quantifies the reconstruction performance available as mask estimation improves.

\subsection{Change outside the masked region}
\label{ssec:exactness}

Compositing operates at native resolution and modifies only masked pixels. Across all 178 samples, the mean absolute change outside the true mask is $0.0$ intensity levels with ground-truth masks and \vSelectedOutside{} levels with predicted masks. The latter value measures the small extension of the predicted mask beyond the annotation, while pixels outside the predicted mask remain unchanged by construction.

The deployed mask covers \vSelectedInstrumentCoverage~per cent of annotated instrument pixels on the validation set, corresponding to recall \vSelectedRecall{}. Precision \vSelectedPrecision{} and IoU \vSelectedIoU{} characterise the predicted-mask boundary, with the corresponding held-out results given in Section~\ref{ssec:seg_results}. False-positive MAE in Table~\ref{tab:ablation} provides the associated pixel-level measurement.

\begin{table*}[pos=t]
\centering\small
\caption{Masked-region fidelity on matched samples. Learned baselines use ground-truth masks. Rows with predicted masks show deployed TransGHOST. Bnd.\ denotes boundary MAE. LPIPS is reported for validation.}
\label{tab:learned_comparison}
\begin{tabular}{llccccc}
\toprule
Method & Masks & PSNR $\uparrow$ & SSIM $\uparrow$ & MAE $\downarrow$ & Bnd.\ $\downarrow$ & LPIPS $\downarrow$\\
\midrule
\multicolumn{7}{l}{\emph{Validation, 178 samples}}\\
LaMa~\cite{suvorov2021resolution}, generative, no ref. & true & 31.89 & 0.891 & 5.29 & 2.00 & 0.0335 \\
Real-time TransRef-inspired~\cite{TransRef}, generative, ref. & true & 29.80 & 0.879 & 6.79 & 2.87 & 0.0548 \\
TransGHOST safe mode, non-generative, ref. & true & \textbf{35.34} & \textbf{0.932} & \textbf{3.48} & \textbf{1.47} & \textbf{0.0206} \\
TransGHOST safe mode, non-generative, ref. & predicted & 31.89 & 0.915 & 4.10 & 2.24 & 0.0377 \\
\midrule
\multicolumn{7}{l}{\emph{Held-out, 101 samples}}\\
LaMa~\cite{suvorov2021resolution}, generative, no ref. & true & 30.17 & 0.866 & 6.46 & 2.21 & - \\
Real-time TransRef-inspired~\cite{TransRef}, generative, ref. & true & 27.54 & 0.847 & 8.41 & 3.51 & - \\
TransGHOST safe mode, non-generative, ref. & true & \textbf{33.94} & \textbf{0.918} & \textbf{4.28} & \textbf{1.72} & - \\
TransGHOST safe mode, non-generative, ref. & predicted & 29.83 & 0.893 & 5.01 & 2.69 & - \\
\bottomrule
\end{tabular}
\end{table*}

\subsection{Comparison with learned inpainters}
\label{ssec:lama}

Table~\ref{tab:learned_comparison} extends the classical controls with reference-free and reference-guided learned methods, evaluated on the same samples with masked scoring. LaMa \cite{suvorov2021resolution} provides the reference-free baseline. The real-time-constrained TransRef-inspired generator from Section~\ref{ssec:inpaint_survey} receives the same reference as TransGHOST and provides a learned reference-guided comparison at its $256$~px operating resolution.

With matched masks, TransGHOST reaches the highest PSNR and SSIM and the lowest MAE, boundary MAE and LPIPS among the compared methods in both evaluation sets. Its PSNR exceeds LaMa by \vLamaMatchedGain~dB on validation (\vOursTruthPSNR{} versus \vLamaValPSNR{}) and by \vLamaMatchedGainHeld~dB on the held-out cohort (\vOursHeldTruthPSNR{} versus \vLamaHeldPSNR{}). Masked MAE is $34$~per cent lower on both sets, and masked SSIM is higher by $0.041$ and $0.052$. Under the masked-crop protocol in Appendix~\ref{app:protocol}, TransGHOST also reduces LPIPS from $0.034$ for LaMa to $0.021$, a $38$~per cent reduction. Because both systems score the same \vSigN{} samples under the same masks, the comparison is paired. The matched-mask differences hold across the cohort rather than resting on a few samples: masked PSNR favours TransGHOST on \vSigMatchedPSNRWins{} of the \vSigN{} samples, masked SSIM on \vSigMatchedSSIMWins{}, masked MAE on \vSigMatchedMAEWins{} and boundary MAE on \vSigMatchedBndWins{}, each separated at $p<0.001$.

The TransRef-inspired generator reaches \vTransrefValPSNR~dB on validation and \vTransrefHeldPSNR~dB on the held-out cohort. Because its training data include the validation recording and two held-out recordings (Section~\ref{ssec:corpus}), these values may overstate its performance. Matched-mask TransGHOST improves these values by \vTransrefMatchedGain{} and \vTransrefMatchedGainHeld~dB, respectively, and also achieves lower MAE, boundary error and LPIPS. Within this real-time-constrained learned comparison, the results support retaining native-resolution geometric transfer in the deployed system.

Mask matching is necessary for a valid restoration comparison because neither learned baseline predicts an instrument mask and both therefore receive the ground-truth mask. Comparing them directly with a TransGHOST configuration that predicts its own mask would confound segmentation and restoration. We therefore report matched-mask results for restoration and place the deployed configuration in a separate, labelled row.

When TransGHOST predicts its own masks, its masked MAE remains $23$~per cent below LaMa on validation and $22$~per cent below it on the held-out cohort, even though LaMa receives ground-truth masks. TransGHOST also has higher masked SSIM on both sets, and its validation masked PSNR equals LaMa's to two decimal places, a difference the paired test does not resolve ($p=\vSigDeployedPSNRP$). Its masked SSIM and MAE advantages are separated at $p<0.001$. Boundary MAE is higher in the mean, at $2.24$ against $2.00$, but the paired test does not resolve that difference either ($p=\vSigDeployedBndP$), and masked LPIPS is higher, at $0.038$ against $0.034$.

\paragraph{Provenance.}
Provenance is a separate distinction that the fidelity columns do not capture. The learned methods infer the content inside the mask, whereas each safe-mode replacement is linked to a recorded frame and source coordinate.

For both learned baselines, the generated region is composited into the original frame using the supplied mask, following LaMa's inference procedure. Table~\ref{tab:learned_comparison} therefore focuses on masked-region fidelity, since the common compositing step fixes their outside-mask change at $0.0$. The synchronised latency of the complete TransGHOST pipeline is reported separately in Section~\ref{ssec:realtime}.

\begin{table*}[pos=t]
\centering\small
\caption{Reference-selection rules on unobstructed and controlled-occlusion candidate sets. Abst.\ denotes the proportion of reconstruction pixels marked as abstentions.}
\label{tab:reference_selection}
\begin{tabular}{lcccccc}
\toprule
& \multicolumn{2}{c}{Unobstructed references} & \multicolumn{4}{c}{Occluded references}\\
\cmidrule(lr){2-3}\cmidrule(lr){4-7}
Selection rule & PSNR & Abst. & PSNR & $\Delta$ & SSIM & Abst.\\
\midrule
Newest valid reference (recency) & 33.42 & 0.000 & 26.44 & - & 0.854 & 0.154 \\
Composite quality score & \emph{n/a} & \emph{n/a} & 27.14 & +0.70 & 0.862 & 0.134 \\
Least region difference~\cite{Hasan2023} & \emph{n/a} & \emph{n/a} & 26.41 & -0.03 & 0.857 & 0.145 \\
Least mask overlap (ours) & 33.37 & 0.000 & \textbf{30.04} & \textbf{+3.60} & 0.904 & 0.037 \\
\bottomrule
\end{tabular}
\end{table*}

\subsection{Reference selection and abstention}
\label{ssec:selection}

We isolate the effect of reference selection on a benchmark in which candidate references may themselves be occluded, a condition under which the overlap rule improves masked PSNR by \vOccGain~dB over recency.

Conventional composited benchmarks obtain a known target by placing an instrument over an instrument-free frame and supplying another instrument-free frame from the same interval as the reference \cite{Hasan2021}. In this setting, every candidate has zero mask overlap with the target region, so the overlap rule reduces to its recency tie-break by design. The occluded-reference benchmark introduces the visibility variation needed to measure the regional term directly.

The occluded-reference benchmark modifies the same samples by placing instruments in a controlled share of the candidate references. Some instruments cover the target region and provide partial support, others appear elsewhere in the frame and leave the target region visible, and the remaining references are unobstructed. This design distinguishes target-region visibility from whole-frame instrument occupancy and also includes cases in which every candidate covers part of the target region. Among 178 samples, the most recent reference is occluded over the target in 78, and in 27 every candidate provides only partial support.

Table~\ref{tab:reference_selection} evaluates the same rules with unobstructed and occluded references. With unobstructed references, the overlap rule reaches \vCleanOverlap~dB masked PSNR and recency reaches \vCleanRecency~dB, so recency is ahead by $0.05$~dB. No candidate hides the target region in this setting, so the regional term has nothing to discriminate: the two rules select the same reference on \vSigCleanIdentical{} of the \vSigN{} samples, and the $0.05$~dB comes from the remaining five.

With occluded references, recency reaches \vOccRecency~dB masked PSNR because it can select a frame in which the target region is hidden. The overlap rule instead reaches \vOccOverlap~dB masked PSNR, a gain of \vOccGain~dB with the alignment stack, masks and samples held fixed.

The two rules choose the same reference on \vSigOccIdentical{} of the \vSigN{} samples and produce identical output there. They differ on the remaining \vSigOccDiffering{}, which is also the number of samples in which the most recent reference is occluded over the target, and the overlap rule gives the higher masked PSNR on \vSigOccOverlapWins{} of them. A two-sided Wilcoxon signed-rank test on the paired per-sample values separates the two rules at $p<0.001$.

The composite quality score provides an intermediate comparison: its whole-frame clean-fraction term improves masked PSNR over recency by \vOccRankedGain~dB, reaching \vOccRanked~dB. The overlap rule reaches \vOccOverlap~dB masked PSNR because it ignores instruments that lie outside the target region.

The published criterion of Hasan et~al.\ \cite{Hasan2023} selects the reference that best matches the current frame in a window around the mask. It reaches \vOccMinSSD~dB masked PSNR, against \vOccRecency~dB for recency, and is \vOccGainOverPublished~dB below the overlap rule. This criterion and the overlap rule differ on \vSigPubDiffering{} samples, and the overlap rule gives the higher masked PSNR on \vSigPubOverlapWins{} of them, separated at $p<0.001$. The published criterion also compares image values over a region for every reference, whereas the overlap rule intersects two masks. In this experiment, its 95th-percentile latency is approximately $1.6\times$ that of the overlap rule.

The selection-timing experiment uses a common pool of six references for every rule, while the live pipeline scans the causal buffer in Algorithm~\ref{alg:transghost}. The values in Table~\ref{tab:reference_selection} therefore provide a controlled comparison between selection rules, and deployment latency is reported separately in Section~\ref{ssec:realtime}.

Appearance scoring uses visible anatomy outside the region being filled, where nearby references are similar. The reference mask instead records target-region visibility directly. Appendix~\ref{app:refpolicy} shows that, for unobstructed references, recency matches oracle pre-alignment selection.

\begin{table*}[pos=t]
\centering\small
\caption{Reference-selection gains under light, medium and heavy distributions of controlled occlusion.}
\label{tab:occlusion_severity}
\begin{tabular}{lccccccc}
\toprule
Distribution & Candidates hiding & Partial-support & \multicolumn{3}{c}{Masked PSNR} & \multicolumn{2}{c}{Abstention}\\
\cmidrule(lr){4-6}\cmidrule(lr){7-8}
& the region & samples & recency & overlap & gain & recency & overlap\\
\midrule
Light & 16\% & 6 & 31.29 & 32.63 & \textbf{+1.34} & 0.043 & 0.009 \\
Medium & 41\% & 27 & 26.44 & 30.04 & \textbf{+3.60} & 0.154 & 0.037 \\
Heavy & 64\% & 53 & 23.36 & 27.76 & \textbf{+4.40} & 0.219 & 0.068 \\
\bottomrule
\end{tabular}
\end{table*}

\paragraph{Effect of occlusion severity.}
We vary three aspects of benchmark severity: the frequency of instrument-bearing references, the probability that an instrument covers the target region, and the frequency with which every candidate provides only partial support. Table~\ref{tab:occlusion_severity} shows consistent gains across this range: \vOccGainLight~dB when $16$~per cent of references hide the region, \vOccGain~dB at $41$~per cent and \vOccGainHeavy~dB at $64$~per cent. Across all three settings, the overlap rule abstains between one-fifth and one-third as often as recency.

\paragraph{Abstention.}
The system abstains when no reference supports a pixel with sufficient confidence. With unobstructed references, abstention covers $0.02$~per cent of the region requiring reconstruction. With occluded references, it rises to \vOccAbstainRecency~per cent under recency and \vOccAbstainOverlap~per cent under the overlap rule.

The benchmark records whether each sample contains a reference that exposes the complete target region, allowing reference-selection and support effects to be separated. In the 27 samples constructed so that every candidate covers part of the target, the overlap rule concentrates \vAbstUnavoidableOverlap~per cent of all abstention and marks, on average, \vAbstUnsatOverlap~per cent of their target region. In the remaining 151 samples, where a fully visible candidate exists, its mean abstention is \vAbstUsableOverlap~per cent with a median of zero, against \vAbstUsableRecency~per cent under recency, which may select an occluded reference.

The overlap rule also improves partial reconstruction in the 27 samples for which every candidate covers part of the target. Choosing the least-occluding reference reaches \vAbstUnsatPSNROverlap~dB masked PSNR, compared with \vAbstUnsatPSNRRecency~dB under recency, while the confidence map identifies the pixels supported by the selected reference.

The confidence threshold is stable over a broad range, with unchanged masked PSNR and abstention from $0.50$ to $0.80$ and only measurement-level variation at $0.90$. The selected value lies within this plateau and away from the transition observed at $0.95$, where abstention is $6.8$~per cent rather than $3.7$~per cent and masked PSNR differs by $0.37$~dB. This stability arises because support is determined mainly by whether the source lies within the image and outside the reference mask, rather than by a continuously varying photometric score.

\begin{table*}[pos=t]
\centering\small
\caption{Held-out ablation on 101 samples from three videos excluded from training, using instrument cutouts from the test split.}
\label{tab:heldout}
\begin{tabular}{lcccccc}
\toprule
Stage & PSNR $\uparrow$ & $\Delta$ & SSIM $\uparrow$ & MAE $\downarrow$ & Bnd.\ MAE $\downarrow$ & FP MAE $\downarrow$\\
\midrule
Global affine registration only & 28.25 & - & 0.861 & 6.76 & 3.83 & 6.63 \\
$+$ residual translation & 29.13 & $+$0.89 & 0.883 & 5.81 & 3.26 & 5.50 \\
$+$ sub-pixel refinement & 29.26 & $+$1.01 & 0.885 & 5.71 & 3.19 & 5.36 \\
$+$ per-block local offsets & \textbf{29.83} & $+$1.58 & \textbf{0.893} & \textbf{5.01} & \textbf{2.69} & \textbf{4.34} \\
\bottomrule
\end{tabular}
\end{table*}

\subsection{Held-out evaluation with training-excluded instrument cutouts}
\label{ssec:heldout}

The configuration was selected using the 178 validation samples and fixed before evaluating the cohort in Table~\ref{tab:heldout}. The held-out cohort was processed as a pre-declared batch without fidelity tuning. After the initial latency measurement, the alignment search was optimised on validation data and evaluated once against pre-declared confirmation criteria described in Appendix~\ref{app:latency}.

The held-out evaluation combines video-level separation with cutout-level separation. The 93 foreground configurations in the cohort are built from 110 distinct instrument cutouts, all drawn from the segmentation test split, and none of them occurs in segmentation training or in the training and validation composites. Every scoring target also differs from its supplied reference.

The ablation follows the same monotonic pattern on the held-out cohort. From the first stage to the last, masked PSNR improves by a total of \vHeldGain~dB, reaching \vHeldPSNR~dB masked PSNR and \vHeldSSIM{} masked SSIM on samples not used for configuration selection.

Mask generalisation is evaluated on the held-out cohort, where recall, precision and IoU are \vHeldRecall{}, \vHeldPrecision{} and \vHeldIoU{}, compared with validation values of \vSelectedRecall{}, \vSelectedPrecision{} and \vSelectedIoU{}.

\paragraph{Per-video gains.}
The held-out cohort spans three videos and eight clips. The masked-PSNR gain is positive for every video: $\vHeldGainVideoFirst$, $\vHeldGainVideoSecond$ and $\vHeldGainVideoThird$~dB on \vHeldSamplesVideoFirst, \vHeldSamplesVideoSecond{} and \vHeldSamplesVideoThird{} samples, respectively. The clip-clustered paired bootstrap interval for this gain is $[\vHeldGainClipLow, \vHeldGainClipHigh]$~dB, and the video-clustered interval is $[\vHeldGainVideoLow, \vHeldGainVideoHigh]$~dB, with both intervals excluding zero.

\begin{figure}[pos=t]
\centering
\includegraphics{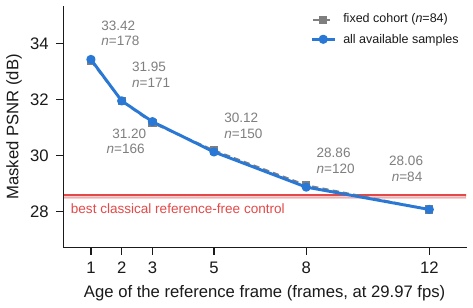}
\caption{Masked fidelity against reference age. The shaded band spans the classical reference-free controls of Table~\ref{tab:ablation}, and the dashed series is the fixed-cohort control of Section~\ref{ssec:lag}.}
\label{fig:reference_age}
\end{figure}

\subsection{Fidelity as a function of reference age}
\label{ssec:lag}

Figure~\ref{fig:reference_age} quantifies reference-guided reconstruction over controlled reference ages. To our knowledge, this temporal fidelity curve has received limited attention in surgical de-occlusion studies.

A one-frame-old reference yields a masked PSNR of \vLagOne~dB, whereas a twelve-frame-old reference yields \vLagTwelve~dB at $0.4$~seconds and 29.97\,fps. This measured interval covers the brief occlusions targeted by TransGHOST, including instrument exchange, and provides an empirical criterion for selecting the reference-age limit in the reported real-time setting. The shaded band in Figure~\ref{fig:reference_age} marks the classical reference-free controls of Table~\ref{tab:ablation}, so the curve locates the reference ages over which geometric transfer exceeds these controls. The comparison with the learned reference-free baseline is a separate measurement and is reported under matched masks in Section~\ref{ssec:lama}.

The available sample count decreases at longer lags because early frames in a clip lack a sufficiently old reference. To separate this change in cohort from the effect of reference age, we repeat the analysis on the \vFixedCohort{} samples available at every lag. The fixed-cohort masked-PSNR curve gives \vLagOneFixed~dB rather than \vLagOne~dB at lag 1 and the same value at lag 12 (Figure~\ref{fig:reference_age}, dashed). A clip-clustered paired bootstrap estimates changes of $\vLagTwoDelta$~dB by lag 2 and $\vLagTwelveDelta$~dB by lag 12, with 95~per cent intervals excluding zero at every lag, as tabulated in Appendix~\ref{app:lag}.

\subsection{Real-time viability}
\label{ssec:realtime}

Real-time operation is assessed here against the interactive criterion adopted in Section~\ref{ssec:latency}: the end-to-end delay between a frame arriving and its restored output must stay within the 100~ms bound reported as acceptable for surgical video feedback in telesurgery and laparoscopy \cite{Nankaku2022, Kumcu2017}. Because occasional slow frames determine whether that bound holds in practice, the criterion is applied to the latency distribution rather than to its mean.

Safe mode has a median latency of \vSelectedPfifty~ms, a validation 95th-percentile latency of \vSelectedPninetyfive~ms, a held-out 95th-percentile latency of \vHeldPninetyfive~ms and a validation 99th percentile of \vSelectedPninetynine~ms on the same hardware. The criterion is met at the median and at the 95th percentile: the validation 95th percentile is approximately $15$~ms below the 100~ms threshold and the median leaves roughly half the budget unused at $1920\times1080$. The 99th percentile lies $1.4$~ms above the threshold, within the $2$ to $3$~ms repeat-measurement spread reported in Appendix~\ref{app:protocol}, so approximately one frame in a hundred reaches the bound. All latency percentiles are measured with the synchronised protocol in Appendix~\ref{app:protocol} (Table~\ref{tab:latency} and Appendix~\ref{app:latency}).

Most latency variation arises from alignment: segmentation latency increases from a median of \vSegMedianMs~ms to \vSegPninetyfiveMs~ms at the 95th percentile, whereas alignment latency increases from a median of \vAlignMedianMs~ms to \vAlignPninetyfiveMs~ms at the 95th percentile and reaches a maximum of \vAlignMaxMs~ms. Alignment cost scales with the warped area, which ranges from approximately $13{,}000$ pixels for a small instrument tip to $1.35$~million when the joint bounding region of two separated instruments covers most of the frame.

The sparse band sampling described in Appendix~\ref{app:latency} reduces the 95th percentile from \vPreBoundPninetyfive{} to \vSelectedPninetyfive~ms and the 99th percentile from \vPreBoundPninetynine{} to \vSelectedPninetynine~ms, with a masked-PSNR change of \vBoundedSearchCost~dB. An additional limit defines a lower-latency operating point with a further \vCeilingSavingMs~ms reduction at the 95th percentile, while the deployed configuration retains the higher-fidelity setting.

\section{Discussion}
\label{sec:discussion}

\subsection{Role in robot-assisted microsurgery}
\label{ssec:discuss_robotics}

Instrument occlusion is a structured failure of intraoperative scene perception. Unlike many image degradations, the missing content may have been visible a fraction of a second earlier, and the affected region is estimated directly by segmenting the instrument. These two conditions make non-generative reconstruction from a recent frame possible.

In a robotic perception stack, provenance allows each downstream component to decide whether reference-derived pixels are admissible. A display may present the full composite, whereas a component requiring direct visual evidence can use the provenance and abstention masks to exclude unsupported pixels. A tracker or registration module can thus retain features beneath the instrument where a recorded frame supports them and treat abstained pixels as missing data, rather than receiving synthesised content that it cannot distinguish from observation. Reporting the median, 95th and 99th percentiles also supports latency budgeting for TransGHOST within perception, monitoring and visual-servoing loops.

The resulting interface combines the reconstructed image with confidence, provenance and abstention layers and gives robotic components visual context together with the source of each reconstructed pixel. Integration with a robotic system and its effect on tracking or control were not evaluated in this study and remain future work. In particular, TransGHOST was evaluated as a perception component on recorded surgical video and was not deployed in a closed-loop robotic manipulation pipeline. A closed-loop evaluation would place it between the microscope camera and the perception components of a surgical robot and compare task performance under instrument occlusion with and without the reconstructed view and its abstention mask. The latency percentiles reported here give the delay that such a loop would have to accommodate.

\subsection{Comparison with existing work}
\label{ssec:discuss_compare}

Xie et~al.\ \cite{Xie2023} provide one relevant clinical comparison among the reference-guided approaches discussed in Section~\ref{sec:related}. Their evaluation uses human endoscopic video, while ours uses a veterinary clinical ophthalmic corpus that permits compositing and masked ground-truth evaluation (Section~\ref{ssec:corpus}). The methods address complementary occlusion regimes: Xie et~al.\ target large, sustained tool-channel occlusions and use a generative term when stored background is insufficient, whereas TransGHOST targets brief, repeated occlusions and preserves pixel provenance in safe mode. The reference-age experiment measures fidelity over this operating range.

The two designs differ in how they achieve coverage. Hybrid reconstruction prioritises a complete image by allowing generated content, while TransGHOST safe mode prioritises traceable support and records abstention explicitly.

Within the established areas of reference-guided surgical de-occlusion \cite{Koreeda2016, Hasan2021} and real-time ophthalmic de-occlusion \cite{Liu2025DALK}, TransGHOST combines a non-generative safe output with causal single-camera operation, per-pixel provenance, measured abstention and evaluation that separates every scoring target from its supplied reference.

\subsection{Operating scope and validation}
\label{ssec:limitations}

The study evaluates TransGHOST using veterinary clinical recordings of canine cataract surgery. The recordings come from manual surgery with hand-held instruments, and occlusion by robot-held instruments is not represented in the corpus. The held-out evaluation uses instrument cutouts excluded from training and reports positive gains for every held-out video (Section~\ref{ssec:heldout}).

Masked-region fidelity is measured on semi-synthetic composites, which provide the clean target required for pixel-level scoring while retaining real surgical backgrounds and instrument cutouts. On unmodified continuous video no such target exists, because the tissue behind the instrument is never recorded in the frame being scored. Restoration fidelity is therefore not measurable on such video, and no fidelity result is claimed for it.

\subsection{Safety and provenance}
\label{ssec:safety}

Safe mode achieves the highest measured fidelity in these experiments, but fidelity and provenance remain separate properties. A generative method could obtain the same score while producing pixels that are inferred rather than traceable to a recorded frame. Provenance must therefore be reported directly rather than inferred from an image-quality metric.

For this reason, each frame is accompanied by per-pixel confidence, an abstention mask and a provenance map. Pixels without valid reference support retain their current-frame values and are marked as abstentions.

\section{Conclusion}
\label{sec:conclusion}

TransGHOST treats brief, repeated instrument occlusion in ophthalmic microsurgery as a reference-guided scene-reconstruction problem. We evaluate this clinically motivated problem using veterinary clinical surgical data. The deployed safe mode uses geometric resampling of recorded image content. The reconstructed image and its confidence, provenance and abstention layers are intended for the perception components of robot-assisted microsurgical systems as well as for the surgeon's display. Evaluation within a closed-loop robotic system remains future work.

Safe mode is causal, uses a single camera and preserves pixels outside the predicted mask, with a mean absolute change of \vSelectedOutside{} intensity levels when measured outside the true mask. Each replacement is linked to its reference frame and source coordinate, while unsupported pixels remain unchanged and are marked as abstentions. The fidelity-versus-reference-age curve further defines the temporal range over which a recent reference remains useful.

On masked-region evaluation, TransGHOST reaches \vSelectedPSNR~dB masked PSNR and \vSelectedSSIM{} masked SSIM at a median end-to-end latency of \vSelectedPfifty~ms and a 95th-percentile latency of \vSelectedPninetyfive~ms. On the held-out cohort, it reaches \vHeldPSNR~dB masked PSNR at \vHeldPninetyfive~ms for the same latency percentile. With masks matched, it exceeds LaMa by \vLamaMatchedGain~dB in masked PSNR and reaches the best value on every reported fidelity metric among the compared methods. When predicting its own masks, it retains lower masked MAE and higher masked SSIM than LaMa while matching validation masked PSNR. When references may be occluded, regional selection improves masked PSNR by \vOccGain~dB over recency.

Upon acceptance, the source code and model architecture will be released in a public GitHub repository, without the dataset. The corpus, evaluation sets and trained weights are available on request under the conditions in the Data availability statement.

\section*{CRediT authorship contribution statement}
\label{sec:credit}

\textbf{Niran Nataraj:} Conceptualization, Methodology, Software, Validation, Formal analysis, Investigation, Data curation, Writing - original draft, Writing - review and editing, Visualization.
\textbf{Maina Sogabe:} Conceptualization, Methodology, Formal analysis, Resources, Writing - original draft, Writing - review and editing, Supervision, Funding acquisition.
\textbf{Norihiko Ito:} Resources, Data curation.
\textbf{Kenji Kawashima:} Formal analysis, Resources, Writing - original draft, Writing - review and editing, Supervision, Project administration.

\section*{Funding}
\label{sec:funding}
This work was supported by JSPS KAKENHI Grant Number 21K18074.

\section*{Declaration of competing interest}
\label{sec:coi}

Kenji Kawashima holds stock in RiverField Inc., a surgical robotics company. Niran Nataraj, Maina Sogabe and Norihiko Ito declare that they have no known competing financial interests that could have appeared to influence the work reported in this paper.

\section*{Data availability}
\label{sec:data_availability}

The source code and model architecture supporting the findings of this study will be released in a public GitHub repository upon acceptance. The release will not include the dataset.

The annotated veterinary clinical image corpus, derived evaluation sets, trained model weights and underlying surgical video recordings are available from the corresponding author on reasonable request, subject to institutional approval and applicable ethical, ownership and data-use restrictions.

\section*{Declaration of generative AI and AI-assisted technologies in the writing process}
\label{sec:genai}

During the preparation of this work, the authors used Grammarly to improve the clarity, readability, and academic presentation of the English text. After using this service, the authors reviewed and edited the manuscript as needed and take full responsibility for the content of the published article.

\section*{Acknowledgements}
\label{sec:ack}
We thank Dr. Masamichi Yamashita of Tottori University for his administrative support.

\bibliographystyle{elsarticle-num}
\bibliography{references}

\begin{thebibliography}{10}
\expandafter\ifx\csname url\endcsname\relax
  \def\url#1{\texttt{#1}}\fi
\expandafter\ifx\csname urlprefix\endcsname\relax\def\urlprefix{URL }\fi
\expandafter\ifx\csname href\endcsname\relax
  \def\href#1#2{#2} \def\path#1{#1}\fi

\bibitem{Ribeiro2022Advances}
L.~Ribeiro, J.~Oliveira, D.~Kuroiwa, M.~Kolko, R.~Fernandes, O.~Junior,
  N.~Moraes, H.~Vasconcelos, T.~Oliveira, M.~Maia, Advances in vitreoretinal
  surgery, Journal of Clinical Medicine 11~(21) (2022) 6428.
\newblock \href {https://doi.org/10.3390/jcm11216428}
  {\path{doi:10.3390/jcm11216428}}.

\bibitem{Eckardt2016}
C.~Eckardt, E.~B. Paulo, Heads-up surgery for vitreoretinal procedures: An
  experimental and clinical study, Retina 36~(1) (2016) 137-147.
\newblock \href {https://doi.org/10.1097/IAE.0000000000000689}
  {\path{doi:10.1097/IAE.0000000000000689}}.

\bibitem{Mukash2025}
G.~Mukash, P.~Vartianen, M.~Torkamani-Azar, Z.~Karadis, M.~Faraz, R.~Bednarik,
  P.~A. Karjalainen, M.~Iso-Mustajärvi, A.~Hussein, Evaluation of visual
  ergonomics in microsurgery: a real-time video processing solution, Acta
  Neurochirurgica 167~(1) (2025) 286.
\newblock \href {https://doi.org/10.1007/s00701-025-06694-2}
  {\path{doi:10.1007/s00701-025-06694-2}}.

\bibitem{Way2003}
L.~W. Way, L.~Stewart, W.~Gantert, K.~Liu, C.~M. Lee, K.~Whang, J.~G. Hunter,
  Causes and prevention of laparoscopic bile duct injuries: Analysis of 252
  cases from a human factors and cognitive psychology perspective, Annals of
  Surgery 237~(4) (2003) 460-469.
\newblock \href {https://doi.org/10.1097/01.SLA.0000060680.92690.E9}
  {\path{doi:10.1097/01.SLA.0000060680.92690.E9}}.

\bibitem{Morita2020}
S.~Morita, H.~Tabuchi, H.~Masumoto, H.~Tanabe, N.~Kamiura, Real-time surgical
  problem detection and instrument tracking in cataract surgery, Journal of
  Clinical Medicine 9~(12) (2020) 3896.
\newblock \href {https://doi.org/10.3390/jcm9123896}
  {\path{doi:10.3390/jcm9123896}}.

\bibitem{Iordachita2022}
I.~I. Iordachita, M.~D. de~Smet, G.~Naus, M.~Mitsuishi, C.~N. Riviere, Robotic
  assistance for intraocular microsurgery: Challenges and perspectives,
  Proceedings of the IEEE 110~(7) (2022) 893--908.
\newblock \href {https://doi.org/10.1109/JPROC.2022.3169466}
  {\path{doi:10.1109/JPROC.2022.3169466}}.

\bibitem{DeBacker2023}
P.~{De Backer}, C.~{Van Praet}, J.~Simoens, M.~{Peraire Lores}, H.~Creemers,
  K.~Mestdagh, C.~Allaeys, S.~Vermijs, P.~Piazza, A.~Mottaran, C.~A. Bravi,
  M.~Paciotti, L.~Sarchi, R.~Farinha, S.~Puliatti, F.~Cisternino, F.~Ferraguti,
  C.~Debbaut, G.~{De Naeyer}, K.~Decaestecker, A.~Mottrie, Improving augmented
  reality through deep learning: Real-time instrument delineation in robotic
  renal surgery, European Urology 84~(1) (2023) 86-91.
\newblock \href {https://doi.org/10.1016/j.eururo.2023.02.024}
  {\path{doi:10.1016/j.eururo.2023.02.024}}.

\bibitem{Hofman2024Deocclusion}
J.~Hofman, P.~{De Backer}, I.~Manghi, J.~Simoens, R.~{De Groote}, H.~{Van Den
  Bossche}, M.~D'Hondt, T.~Oosterlinck, J.~Lippens, C.~{Van Praet},
  F.~Ferraguti, C.~Debbaut, Z.~Li, O.~Kutter, A.~Mottrie, K.~Decaestecker,
  First-in-human real-time {AI}-assisted instrument deocclusion during
  augmented reality robotic surgery, Healthcare Technology Letters 11~(2-3)
  (2024) 33-39.
\newblock \href {https://doi.org/10.1049/htl2.12056}
  {\path{doi:10.1049/htl2.12056}}.

\bibitem{Edwards2018}
T.~L. Edwards, K.~Xue, H.~C.~M. Meenink, M.~J. Beelen, G.~J.~L. Naus, M.~P.
  Simunovic, M.~Latasiewicz, A.~D. Farmery, M.~D. de~Smet, R.~E. MacLaren,
  First-in-human study of the safety and viability of intraocular robotic
  surgery, Nature Biomedical Engineering 2~(9) (2018) 649--656.
\newblock \href {https://doi.org/10.1038/s41551-018-0248-4}
  {\path{doi:10.1038/s41551-018-0248-4}}.

\bibitem{Becker2013}
B.~C. Becker, R.~A. MacLachlan, L.~A. Lobes, G.~D. Hager, C.~N. Riviere,
  Vision-based control of a handheld surgical micromanipulator with virtual
  fixtures, IEEE Transactions on Robotics 29~(3) (2013) 674--683.
\newblock \href {https://doi.org/10.1109/TRO.2013.2239552}
  {\path{doi:10.1109/TRO.2013.2239552}}.

\bibitem{Kim2020}
J.~W. Kim, C.~He, M.~Urias, P.~Gehlbach, G.~D. Hager, I.~Iordachita,
  M.~Kobilarov, Autonomously navigating a surgical tool inside the eye by
  learning from demonstration, in: 2020 IEEE International Conference on
  Robotics and Automation (ICRA), 2020, pp. 7351--7357.
\newblock \href {https://doi.org/10.1109/ICRA40945.2020.9196537}
  {\path{doi:10.1109/ICRA40945.2020.9196537}}.

\bibitem{Chen2018IRISS}
C.-W. Chen, Y.-H. Lee, M.~J. Gerber, H.~Cheng, Y.-C. Yang, A.~Govetto, A.~A.
  Francone, S.~Soatto, W.~S. Grundfest, J.-P. Hubschman, T.-C. Tsao,
  Intraocular robotic interventional surgical system ({IRISS}): Semi-automated
  {OCT}-guided cataract removal, The International Journal of Medical Robotics
  and Computer Assisted Surgery 14~(6) (2018) e1949.
\newblock \href {https://doi.org/10.1002/rcs.1949}
  {\path{doi:10.1002/rcs.1949}}.

\bibitem{Braun2018EyeSLAM}
D.~Braun, S.~Yang, J.~N. Martel, C.~N. Riviere, B.~C. Becker, {EyeSLAM}:
  Real-time simultaneous localization and mapping of retinal vessels during
  intraocular microsurgery, The International Journal of Medical Robotics and
  Computer Assisted Surgery 14~(1) (2018) e1848.
\newblock \href {https://doi.org/10.1002/rcs.1848}
  {\path{doi:10.1002/rcs.1848}}.

\bibitem{Long2021EDSSR}
Y.~Long, Z.~Li, C.~H. Yee, C.~F. Ng, R.~H. Taylor, M.~Unberath, Q.~Dou,
  {E-DSSR}: Efficient dynamic surgical scene reconstruction with
  transformer-based stereoscopic depth perception, in: Medical Image Computing
  and Computer Assisted Intervention -- MICCAI 2021, 2021, pp. 415--425.
\newblock \href {https://doi.org/10.1007/978-3-030-87202-1_40}
  {\path{doi:10.1007/978-3-030-87202-1_40}}.

\bibitem{Wang2022EndoNeRF}
Y.~Wang, Y.~Long, S.~H. Fan, Q.~Dou, Neural rendering for stereo {3D}
  reconstruction of deformable tissues in robotic surgery, in: Medical Image
  Computing and Computer Assisted Intervention -- MICCAI 2022, 2022, pp.
  431--441.
\newblock \href {https://doi.org/10.1007/978-3-031-16449-1_41}
  {\path{doi:10.1007/978-3-031-16449-1_41}}.

\bibitem{Kovacs2021}
G.~Kovács, A.~Fazekas, A new baseline for retinal vessel segmentation:
  Numerical identification and correction of methodological inconsistencies
  affecting 100+ papers, Medical Image Analysis 75 (2022) 102300.
\newblock \href {https://doi.org/10.1016/j.media.2021.102300}
  {\path{doi:10.1016/j.media.2021.102300}}.

\bibitem{Hatamizadeh2022}
A.~Hatamizadeh, H.~Hosseini, N.~Patel, J.~Choi, C.~C. Pole, C.~M. Hoeferlin,
  S.~D. Schwartz, D.~Terzopoulos, {RAVIR}: A dataset and methodology for the
  semantic segmentation and quantitative analysis of retinal arteries and veins
  in infrared reflectance imaging, IEEE Journal of Biomedical and Health
  Informatics 26~(7) (2022) 3272-3283.
\newblock \href {https://doi.org/10.1109/JBHI.2022.3163352}
  {\path{doi:10.1109/JBHI.2022.3163352}}.

\bibitem{Nespolo2022}
R.~G. Nespolo, D.~Yi, E.~Cole, D.~Wang, A.~Warren, Y.~I. Leiderman, Feature
  tracking and segmentation in real time via deep learning in vitreoretinal
  surgery: A platform for artificial intelligence-mediated surgical guidance,
  Ophthalmology Retina 7~(3) (2023) 236-242.
\newblock \href {https://doi.org/10.1016/j.oret.2022.10.002}
  {\path{doi:10.1016/j.oret.2022.10.002}}.

\bibitem{Sogabe2022Detection}
M.~Sogabe, N.~Ito, T.~Miyazaki, T.~Kawase, T.~Kanno, K.~Kawashima, Detection of
  instruments inserted into eye in cataract surgery using single-shot multibox
  detector, Sensors and Materials 34~(1) (2022) 47-54.
\newblock \href {https://doi.org/10.18494/SAM3762}
  {\path{doi:10.18494/SAM3762}}.

\bibitem{Yeh2023PhacoTrainer}
H.-H. Yeh, A.~M. Jain, O.~Fox, K.~Sebov, S.~Y. Wang, {PhacoTrainer}: Deep
  learning for cataract surgical videos to track surgical tools, Translational
  Vision Science \& Technology 12~(3) (2023) 23.
\newblock \href {https://doi.org/10.1167/tvst.12.3.23}
  {\path{doi:10.1167/tvst.12.3.23}}.

\bibitem{Borkovkina2020}
S.~Borkovkina, A.~Camino, W.~Janpongsri, M.~V. Sarunic, Y.~Jian, Real-time
  retinal layer segmentation of {OCT} volumes with {GPU}-accelerated
  inferencing using a compressed, low-latency neural network, Biomedical Optics
  Express 11~(7) (2020) 3968-3984.
\newblock \href {https://doi.org/10.1364/BOE.395279}
  {\path{doi:10.1364/BOE.395279}}.

\bibitem{Huang2023}
Y.~Huang, R.~Asaria, D.~Stoyanov, M.~Sarunic, S.~Bano, {PseudoSegRT}: Efficient
  pseudo-labelling for intraoperative {OCT} segmentation, International Journal
  of Computer Assisted Radiology and Surgery 18~(7) (2023) 1245-1252.
\newblock \href {https://doi.org/10.1007/s11548-023-02928-9}
  {\path{doi:10.1007/s11548-023-02928-9}}.

\bibitem{fgvc20}
C.~Gao, A.~Saraf, J.-B. Huang, J.~Kopf, Flow-edge guided video completion, in:
  Proceedings of the European Conference on Computer Vision (ECCV), 2020, pp.
  713-729.
\newblock \href {https://doi.org/10.1007/978-3-030-58610-2_42}
  {\path{doi:10.1007/978-3-030-58610-2_42}}.

\bibitem{e2fgvi22}
Z.~Li, C.-Z. Lu, J.~Qin, C.-L. Guo, M.-M. Cheng, Towards an end-to-end
  framework for flow-guided video inpainting, in: Proceedings of the IEEE/CVF
  Conference on Computer Vision and Pattern Recognition (CVPR), 2022, pp.
  17541-17550.
\newblock \href {https://doi.org/10.1109/CVPR52688.2022.01704}
  {\path{doi:10.1109/CVPR52688.2022.01704}}.

\bibitem{zhou2023pro}
S.~Zhou, C.~Li, K.~C.~K. Chan, C.~C. Loy, {ProPainter}: Improving propagation
  and transformer for video inpainting, in: Proceedings of the IEEE/CVF
  International Conference on Computer Vision (ICCV), 2023, pp. 10443-10452.
\newblock \href {https://doi.org/10.1109/ICCV51070.2023.00961}
  {\path{doi:10.1109/ICCV51070.2023.00961}}.

\bibitem{Koreeda2016}
Y.~Koreeda, Y.~Kobayashi, S.~Ieiri, Y.~Nishio, K.~Kawamura, S.~Obata,
  R.~Souzaki, M.~Hashizume, M.~G. Fujie, Virtually transparent surgical
  instruments in endoscopic surgery with augmentation of obscured regions,
  International Journal of Computer Assisted Radiology and Surgery 11~(10)
  (2016) 1927-1936.
\newblock \href {https://doi.org/10.1007/s11548-016-1384-5}
  {\path{doi:10.1007/s11548-016-1384-5}}.

\bibitem{Hasan2021}
S.~M.~K. Hasan, R.~A. Simon, C.~A. Linte, Segmentation and removal of surgical
  instruments for background scene visualization from endoscopic/laparoscopic
  video, in: Medical Imaging 2021: Image-Guided Procedures, Robotic
  Interventions, and Modeling, Vol. 11598, 2021, p. 115980A.
\newblock \href {https://doi.org/10.1117/12.2580668}
  {\path{doi:10.1117/12.2580668}}.

\bibitem{Hasan2023}
S.~M.~K. Hasan, R.~A. Simon, C.~A. Linte, Inpainting surgical occlusion from
  laparoscopic video sequences for robot-assisted interventions, Journal of
  Medical Imaging 10~(4) (2023) 045002.
\newblock \href {https://doi.org/10.1117/1.JMI.10.4.045002}
  {\path{doi:10.1117/1.JMI.10.4.045002}}.

\bibitem{Li2024optimised}
P.~Li, L.~Liu, C.-B. Sch{\"o}nlieb, A.~I. Aviles-Rivero, Optimised {ProPainter}
  for video diminished reality inpainting, in: 2024 IEEE International
  Symposium on Biomedical Imaging (ISBI), 2024, pp. 1-5.
\newblock \href {https://doi.org/10.1109/ISBI56570.2024.10635380}
  {\path{doi:10.1109/ISBI56570.2024.10635380}}.

\bibitem{Xie2023}
D.~Xie, W.~Chen, J.~Zhao, X.~Song, K.~Wang, W.~Xia, H.~Liu, F.~Chang, C.~Li,
  X.~Duan, Surgical instruments hyalinization: Occlusion removal in minimally
  invasive endoscopic surgery, Biomimetic Intelligence and Robotics 3~(3)
  (2023) 100105.
\newblock \href {https://doi.org/10.1016/j.birob.2023.100105}
  {\path{doi:10.1016/j.birob.2023.100105}}.

\bibitem{Celdran2025}
F.~J. Celdrán, J.~Jiménez-Ruescas, C.~Lobato, L.~Salazar, J.~A.
  Sánchez-Margallo, F.~M. Sánchez-Margallo, P.~González, Use of augmented
  reality for training assistance in laparoscopic surgery: Scoping literature
  review, Journal of Medical Internet Research 27 (2025) e58108.
\newblock \href {https://doi.org/10.2196/58108} {\path{doi:10.2196/58108}}.

\bibitem{Nankaku2022}
A.~Nankaku, M.~Tokunaga, H.~Yonezawa, T.~Kanno, K.~Kawashima, K.~Hakamada,
  S.~Hirano, E.~Oki, M.~Mori, Y.~Kinugasa, Maximum acceptable communication
  delay for the realization of telesurgery, PLOS ONE 17~(10) (2022) e0274328.
\newblock \href {https://doi.org/10.1371/journal.pone.0274328}
  {\path{doi:10.1371/journal.pone.0274328}}.

\bibitem{Kumcu2017}
A.~Kumcu, L.~Vermeulen, S.~A. Elprama, P.~Duysburgh, L.~Platiša, Y.~{Van
  Nieuwenhove}, N.~{Van De Winkel}, A.~Jacobs, J.~{Van Looy}, W.~Philips,
  Effect of video lag on laparoscopic surgery: Correlation between performance
  and usability at low latencies, The International Journal of Medical Robotics
  and Computer Assisted Surgery 13~(2) (2017) e1758.
\newblock \href {https://doi.org/10.1002/rcs.1758}
  {\path{doi:10.1002/rcs.1758}}.

\bibitem{Liu2025DALK}
W.~Liu, J.~Pan, L.~Jia, S.~Rao, J.~Zang, Real-time tracking and inpainting
  network with joint learning iterative modules for {AR}-based {DALK} surgical
  navigation, Computer Methods and Programs in Biomedicine 272 (2025) 109068.
\newblock \href {https://doi.org/10.1016/j.cmpb.2025.109068}
  {\path{doi:10.1016/j.cmpb.2025.109068}}.

\bibitem{Conze2016}
P.-H. Conze, P.~Robert, T.~Crivelli, L.~Morin, Multi-reference combinatorial
  strategy towards longer long-term dense motion estimation, Computer Vision
  and Image Understanding 150 (2016) 66-80.
\newblock \href {https://doi.org/10.1016/j.cviu.2016.04.013}
  {\path{doi:10.1016/j.cviu.2016.04.013}}.

\bibitem{Li2020optical}
B.~Li, J.~Han, Y.~Xu, K.~Rose, Optical flow based co-located reference frame
  for video compression, IEEE Transactions on Image Processing 29 (2020)
  8303-8315.
\newblock \href {https://doi.org/10.1109/TIP.2020.3014723}
  {\path{doi:10.1109/TIP.2020.3014723}}.

\bibitem{CaDIS}
M.~Grammatikopoulou, E.~Flouty, A.~Kadkhodamohammadi, G.~Quellec, A.~Chow,
  J.~Nehme, I.~Luengo, D.~Stoyanov, {CaDIS}: Cataract dataset for surgical
  {RGB}-image segmentation, Medical Image Analysis 71 (2021) 102053.
\newblock \href {https://doi.org/10.1016/j.media.2021.102053}
  {\path{doi:10.1016/j.media.2021.102053}}.

\bibitem{Ni2019Cata7}
Z.-L. Ni, G.-B. Bian, X.-H. Zhou, Z.-G. Hou, X.-L. Xie, C.~Wang, Y.-J. Zhou,
  R.-Q. Li, Z.~Li, {RAUNet}: Residual attention {U-Net} for semantic
  segmentation of cataract surgical instruments, in: Neural Information
  Processing (ICONIP), 2019, pp. 139-149.
\newblock \href {https://doi.org/10.1007/978-3-030-36711-4_13}
  {\path{doi:10.1007/978-3-030-36711-4_13}}.

\bibitem{suvorov2021resolution}
R.~Suvorov, E.~Logacheva, A.~Mashikhin, A.~Remizova, A.~Ashukha, A.~Silvestrov,
  N.~Kong, H.~Goka, K.~Park, V.~Lempitsky, Resolution-robust large mask
  inpainting with {F}ourier convolutions, in: Proceedings of the IEEE/CVF
  Winter Conference on Applications of Computer Vision (WACV), 2022, pp.
  3172-3182.
\newblock \href {https://doi.org/10.1109/WACV51458.2022.00323}
  {\path{doi:10.1109/WACV51458.2022.00323}}.

\bibitem{Ouyang2021Internal}
H.~Ouyang, T.~Wang, Q.~Chen, Internal video inpainting by implicit long-range
  propagation, in: Proceedings of the IEEE/CVF International Conference on
  Computer Vision (ICCV), 2021, pp. 14559-14568.
\newblock \href {https://doi.org/10.1109/ICCV48922.2021.01431}
  {\path{doi:10.1109/ICCV48922.2021.01431}}.

\bibitem{TransRef}
T.~Liu, L.~Liao, D.~Chen, J.~Xiao, Z.~Wang, C.-W. Lin, S.~Satoh, {TransRef}:
  Multi-scale reference embedding transformer for reference-guided image
  inpainting, Neurocomputing 632 (2025) 129749.
\newblock \href {https://doi.org/10.1016/j.neucom.2025.129749}
  {\path{doi:10.1016/j.neucom.2025.129749}}.

\bibitem{RGTSI}
T.~Liu, L.~Liao, Z.~Wang, S.~Satoh, Reference-guided texture and structure
  inference for image inpainting, in: 2022 IEEE International Conference on
  Image Processing (ICIP), 2022, pp. 1996-2000.
\newblock \href {https://doi.org/10.1109/ICIP46576.2022.9897592}
  {\path{doi:10.1109/ICIP46576.2022.9897592}}.

\end{thebibliography}

\clearpage
\appendix
\counterwithin*{table}{section}
\counterwithin*{figure}{section}
\renewcommand{\thetable}{\Alph{section}\arabic{table}}
\renewcommand{\thefigure}{\Alph{section}\arabic{figure}}

\section{Evaluation of optional learned refinement}
\label{app:quality}

We evaluated a compact U-Net with windowed reference attention as an optional refinement of the safe composite. The network operates on a $256$~px region of interest around the mask and is trained on the triplets from Section~\ref{ssec:triplets}. Its result is returned as a separate image layer with every altered pixel marked as synthesised.

The comparison uses the 178 validation samples with ground-truth masks in this $256$~px refinement configuration, rather than the native-resolution setting of Table~\ref{tab:ablation}. The unrefined safe composite reaches \vQualityControlPSNR~dB masked PSNR and the learned refinement reaches \vQualityPSNR~dB. The learned variant has a mean end-to-end latency of \vQualityLatencyMean~ms and a 95th percentile of \vQualityPninetyfive~ms on the RTX 5000 Ada used for all timing experiments. Because the refinement was trained on triplets that include the validation recording (Section~\ref{ssec:corpus}), this comparison may favour it. The safe composite gives the higher measured fidelity and keeps direct source traceability, which supports its selection as the reported TransGHOST output.

\section{Reference selection among unobstructed references}
\label{app:refpolicy}

Table~\ref{tab:reference_policy} varies the reference choice while holding the alignment stack, masks and samples fixed. The best and worst valid choices differ by \vSelectionSpread~dB in masked PSNR. Recency reaches the highest masked PSNR among the evaluated policies, including the oracle that selects on highest pre-alignment agreement, which is $0.07$~dB lower. Pre-alignment agreement is a proxy rather than the scored quantity, so a policy that maximises it need not maximise masked PSNR after alignment. When all available references are unobstructed, the measurements therefore favour the newest valid reference.

\begin{table}[pos=t]
\centering\scriptsize
\caption{Reference choice with alignment, masks and samples held fixed.}
\label{tab:reference_policy}
\begin{tabular}{@{}l@{\hspace{10pt}}c@{\hspace{10pt}}c@{\hspace{10pt}}c@{\hspace{10pt}}c@{}}
\toprule
Selection policy & PSNR $\uparrow$ & $\Delta$ & SSIM $\uparrow$ & MAE $\downarrow$\\
\midrule
Nearest valid reference (ours) & \textbf{33.42} & - & \textbf{0.933} & \textbf{2.93} \\
Random valid reference & 31.06 & -2.36 & 0.910 & 4.64 \\
Oldest valid reference & 29.19 & -4.23 & 0.888 & 6.37 \\
Oracle on pre-alignment agreement & 33.35 & -0.07 & 0.931 & 3.02 \\
\bottomrule
\end{tabular}
\end{table}

Every reference in Table~\ref{tab:reference_policy} is unobstructed, so mask overlap is zero and the regional criterion reduces to recency. Section~\ref{ssec:selection} evaluates the overlap term separately on references that may contain an instrument over the target region.

With unobstructed references, the regional criterion reduces to recency by design and avoids the lower fidelity of older or random choices. When references may occlude the target region, the regional term becomes active and improves on recency by \vOccGain~dB.

\section{Evaluation of alternative configurations}
\label{app:rejected}

Table~\ref{tab:alternatives} reports alternatives measured before the latency changes in Appendix~\ref{app:latency}. Their latency values can therefore be compared with one another but are higher than the deployed percentiles in Section~\ref{ssec:realtime}. The later change alters masked PSNR by \vBoundedSearchCost~dB. This table and Appendix~\ref{app:latency} record the pre-change configuration in separate timing sessions, which is why its 95th percentile appears here as $99$~ms and there as \vPreBoundPninetyfive~ms, a difference within the repeat spread stated in Appendix~\ref{app:protocol}.

\begin{table*}[pos=t]
\centering\small
\caption{Design-selection experiments comparing the reported configuration with measured alternatives.}
\label{tab:alternatives}
\begin{tabular}{lcccc}
\toprule
Variant & PSNR $\uparrow$ & $\Delta$ & p95 (ms) $\downarrow$ & p99 (ms) $\downarrow$\\
\midrule
\multicolumn{5}{l}{\emph{Multi-reference evaluation set, one reference per candidate lag}}\\
Single nearest reference (ours) & \textbf{33.42} & - & \textbf{96} & \textbf{120} \\
Median fusion of 5 references & 33.39 & $-0.03$ & 109 & 744 \\
Per-block choice among 5 references & \textbf{33.42} & $0.00$ & 109 & 343 \\
\midrule
\multicolumn{5}{l}{\emph{Base validation set, deployed configuration}}\\
Deployed configuration (ours) & \textbf{31.89} & - & \textbf{99} & \textbf{118} \\
$+$ photometric consistency gate & 31.14 & $-0.75$ & 110 & 122 \\
$+$ alpha unmixing at the boundary & 31.81 & $-0.08$ & 245 & 266 \\
\bottomrule
\end{tabular}
\end{table*}

The nearest single reference matches or exceeds median fusion and per-block choice among five aligned references and has the lowest 99th-percentile latency. After registration, the remaining correspondence error is systematic rather than independent across references, so averaging adds little information. The same design-selection experiment favours direct geometric compositing over a photometric consistency gate and over alpha unmixing at the boundary. The gate can suppress valid reconstruction at short lags, and alpha unmixing depends on additional alpha and instrument-colour estimates and has approximately three times the latency.

Local correspondence fitted to visible pixels outside the mask outperforms dense optical flow within the mask by $0.3$ to $1.1$~dB across the evaluated smoothing settings. The visible-region formulation avoids following instrument motion in the current frame and supports the local-offset design used in Table~\ref{tab:ablation}.

\paragraph{Effect of block granularity.}
With access to ground truth, an oracle choice of translation per block improves masked PSNR by $+1.5$ to $+2.9$~dB as the block size decreases from 64 to 16 pixels. This oracle gain indicates a potential benefit from finer local correspondence. We evaluated practical block sizes of 16, 24, 32 and 48 pixels at search resolutions of 0.35, 0.50 and 0.70 of native resolution, covering eight of the twelve combinations. The deployed setting achieves the highest measured fidelity at \vSelectedPSNR~dB, compared with \vBlockSweepMin{} to \vBlockSweepMax~dB across the alternatives, while also avoiding the higher latency of finer settings. At block sizes below 48 pixels, the visible support becomes too sparse for stable matching.

\section{Latency engineering}
\label{app:latency}

Alignment cost scales with the warped area, which varies by two orders of magnitude: the median is $140{,}000$ pixels, the 95th percentile is $754{,}000$, and the maximum is $1.35$~million. Because the largest regions determine the latency tail, we apply two changes to reduce their computational cost.

The residual-translation search uses only a band of visible pixels outside the mask. The original implementation nevertheless warped the complete region at each stage to obtain these samples. We instead sample the reference directly at the band coordinates and evaluate all offsets in one batched interpolation, so that cost is proportional to the number of band pixels rather than to the full region area. Direct sampling also evaluates the offset in the same reference-coordinate system in which it is ultimately applied.

The blockwise search compares a decimated region for every offset. With fractional decimation alone, large regions remain far more expensive than small ones, so an absolute limit is also placed on the number of pixels compared per offset. Regions below this limit remain unchanged, while only the largest are further coarsened after the global sub-pixel stage has estimated fine displacement.

\begin{figure}[pos=t]
\centering
\includegraphics{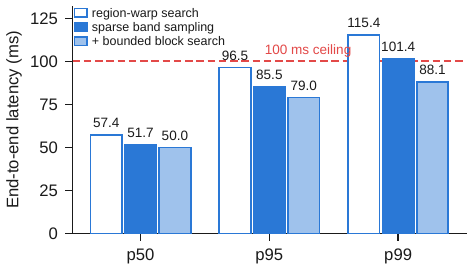}
\caption{End-to-end latency percentiles for three implementations of the alignment search on the validation set. The median changes by several milliseconds, whereas the 99th percentile decreases by approximately one quarter.}
\label{fig:latency}
\end{figure}

Figure~\ref{fig:latency} shows that, on the validation set, sparse band sampling reduces the 95th-percentile end-to-end latency from \vPreBoundPninetyfive~ms to \vSelectedPninetyfive~ms and the 99th percentile from \vPreBoundPninetynine{} to \vSelectedPninetynine~ms, with a masked-PSNR change of \vBoundedSearchCost~dB. An additional absolute pixel limit provides a second operating point with a further \vCeilingSavingMs~ms reduction at the 95th percentile, at approximately \vCeilingCost~dB lower masked PSNR. The reported default retains the higher-fidelity configuration.

\begin{table*}[pos=t]
\centering\small
\caption{Initial and confirmatory end-to-end latency on the held-out cohort.}
\label{tab:latency}
\begin{tabular}{lcccc}
\toprule
Configuration & PSNR $\uparrow$ & p50 (ms) $\downarrow$ & p95 (ms) $\downarrow$ & p99 (ms) $\downarrow$\\
\midrule
\multicolumn{5}{l}{\emph{Initial measurement}}\\
Without per-block local offsets & 29.25 & 57.6 & 87.2 & 100.5 \\
Full configuration & 29.83 & 61.0 & 102.7 & 116.9 \\
\midrule
\multicolumn{5}{l}{\emph{Confirmatory measurement, sparse band sampling}}\\
Without per-block local offsets & 29.26 & 52.1 & 83.1 & 90.0 \\
Full configuration & 29.83 & 55.9 & 86.1 & 101.2 \\
\bottomrule
\end{tabular}
\end{table*}

Table~\ref{tab:latency} reports the initial and confirmatory measurements on the held-out cohort. The initial full configuration reached \vHeldPSNR~dB masked PSNR at $102.7$~ms for the 95th percentile, while the reduced configuration met the latency target at $29.25$~dB masked PSNR. This comparison motivated the latency analysis and the resulting sparse-search changes, which were designed and selected on the validation set with a masked-PSNR effect of \vBoundedSearchCost~dB.

Before the confirmatory held-out evaluation, we specified three acceptance conditions: masked PSNR within $0.10$~dB of the initial result, a 95th percentile below $100$~ms, and unchanged mask metrics because segmentation was not modified. All three conditions were met: the full configuration reaches \vHeldPSNR~dB masked PSNR at \vHeldPninetyfive~ms for this percentile, with masked PSNR unchanged to two decimal places. The reduced alternative decreases latency by $3$~ms and reaches $29.26$~dB masked PSNR.

\section{Fidelity against reference age}
\label{app:lag}

\begin{table*}[pos=t]
\centering\small
\caption{Reference-age analysis at 29.97\,fps. The fixed-cohort block contains samples available at every lag. Bracketed values are 95 per cent intervals from a clip-clustered paired bootstrap against lag 1.}
\label{tab:reference_age}
\begin{tabular}{cccccccc}
\toprule
& \multicolumn{5}{c}{All available samples} & \multicolumn{2}{c}{Fixed cohort}\\
\cmidrule(lr){2-6}\cmidrule(lr){7-8}
Lag & $n$ & PSNR $\uparrow$ & SSIM $\uparrow$ & MAE $\downarrow$ & Bnd.\ MAE $\downarrow$ & PSNR $\uparrow$ & $\Delta$ vs lag 1 [95\% CI]\\
\midrule
1 & 178 & 33.42 & 0.933 & 2.93 & 1.58 & 33.37 & - \\
2 & 171 & 31.95 & 0.920 & 3.75 & 2.04 & 31.94 & $-1.43\;[-2.93, -1.21]$ \\
3 & 166 & 31.20 & 0.910 & 4.35 & 2.38 & 31.16 & $-2.22\;[-4.75, -1.95]$ \\
5 & 150 & 30.12 & 0.896 & 5.47 & 2.96 & 30.17 & $-3.20\;[-7.02, -3.04]$ \\
8 & 120 & 28.86 & 0.884 & 6.85 & 3.65 & 28.93 & $-4.45\;[-7.89, -3.96]$ \\
12 & 84 & 28.06 & 0.883 & 7.36 & 3.99 & 28.06 & $-5.31\;[-9.46, -4.94]$ \\
\bottomrule
\end{tabular}
\end{table*}

Table~\ref{tab:reference_age} reports the numerical values underlying the curve in Section~\ref{ssec:lag}, including the sample count at each lag and the fixed-cohort control that separates reference age from the changing cohort size.

\section{Evaluation protocol details}
\label{app:protocol}

\paragraph{Rationale for masked scoring.} Compositing replaces only masked pixels and leaves the remainder of the frame unchanged. Because masks occupy only a few per cent of the frame, whole-frame metrics are dominated by pixels that no method alters. All fidelity results in Section~\ref{sec:results} are therefore computed over the instrument mask. Masked LPIPS uses the mask-centred crop defined below.

\paragraph{Rationale for a composited benchmark.} Restoration fidelity requires a known view of the tissue behind an instrument. We therefore place a real instrument cutout over an instrument-free frame from a held-out clip and use the unobstructed frame as the scoring target. The supplied reference must be a different image, and samples that violate this condition are excluded before evaluation. This controlled semi-synthetic construction combines real surgical imagery with an exact pixel-level target.

\paragraph{Metric definitions.} Masked PSNR, SSIM and MAE are computed over the true instrument mask. For LPIPS, the mask bounding box is padded by $64$~pixels, clamped to the frame and resized to $256\times256$ before evaluation with LPIPS(VGG). Including context outside the mask allows the metric to capture seams at the restoration boundary. Boundary MAE uses a narrow band around the mask edge, while false-positive MAE reports reconstruction error at pixels included by the predicted mask but absent from the true mask.

\paragraph{Learned baseline implementation.} LaMa uses the released Big-LaMa generator weights at a maximum image side of $1920$~pixels. The real-time-constrained TransRef-inspired generator uses the checkpoint fine-tuned in this project, at its $256$~px operating resolution. It receives the same reference and ground-truth mask as TransGHOST, and its prediction is resized to the native frame before masked compositing. Both learned baselines are evaluated in inference mode on the same samples and with the same masked metrics. Table~\ref{tab:learned_comparison} reports this matched fidelity comparison, while Section~\ref{ssec:realtime} reports the latency of the complete TransGHOST pipeline, measured with device synchronisation.

\paragraph{Latency protocol.} Wall-clock timings are measured on one otherwise-idle RTX 5000 Ada at native $1920\times1080$ resolution. Eight warm-up passes are discarded, and the device is synchronised before and after each measured frame. Repeated measurements vary by approximately $2$ to $3$~ms at the 95th percentile, which defines the resolution used to interpret latency differences.

\paragraph{Reproducibility contact.} Questions about the protocol or requests concerning the evaluation sets may be addressed to Niran Nataraj (niran@g.ecc.u-tokyo.ac.jp) or Maina Sogabe (fsogabe-vet@g.ecc.u-tokyo.ac.jp).

\end{document}